\documentclass[%
preprint,
amsmath,amssymb,
]{revtex4-2}
\usepackage{setspace}%
\usepackage{graphicx}
\usepackage{svg}
\usepackage{subcaption}
\usepackage{dcolumn}
\usepackage{bm}
\usepackage[outdir=./]{epstopdf}

\begin{document}


\title{Interaction between a Rising Bubble and a Stationary Droplet Immersed in a Liquid Pool using Ternary Conservative Phase-Field Lattice Boltzmann Method}

\author{Chunheng Zhao}
\author{Taehun Lee}%
 \email{thlee@ccny.cuny.edu}
\affiliation{%
Department of Mechanical Engineering, City College of New York, New York 10031, USA
}%

\date{\today}

\begin{abstract}
When a stationary bubble and a stationary droplet immersed in a liquid pool are brought in contact with each other, they form a bubble-droplet aggregate. Its equilibrium morphology and stability largely depend on the combination of different components' surface tensions, known as spreading factor. In this study, we look at the interaction between a rising bubble and a stationary droplet to better understand the dynamics of coalescence and rising as well as morphological changes for the bubble-droplet aggregate. A systematic study is conducted on the interaction processes with various bubble sizes and spreading factors. The current simulation framework consists of the ternary conservative phase-field Lattice Boltzmann method (LBM) for interface tracking and the velocity-pressure LBM for hydrodynamics, which is validated for the benchmark cases such as liquid lens and parasitic currents around a static droplet with several popular surface tension formulations. We further test our LBM for the morphology changes of two droplets initially in contact with various spreading factors and depict the final morphologies in a phase diagram. The separated, partially engulfed and completely engulfed morphologies can be replicated by systematically altering the sign of the spreading factors. The rising bubble and stationary droplet interaction is simulated based on the final morphologies obtained under stationary conditions by imposing an imaginary buoyancy force on the rising bubble. The results indicate that the bubble-droplet aggregate with double emulsion morphology can minimize the distotion of the bubble-droplet aggregate and achieve a greater terminal velocity than the aggregate with partially engulfed morphology.
\end{abstract}

\maketitle

\newpage

\section{Introduction}
The rising bubble and droplet interaction is one of the common phenomena found in gas flotation, water cleanup and oil extraction \cite{moosai2003gas,saththasivam2016overview,grattoni2003photographic}. The entire interaction can be divided into following three parts: bubble rising process, bubble droplet interaction process, and aggregate rising process. These processes have been studied to develop an optimal system by experiments and simulations for decades \cite{amaya2010single, amaya2011numerical,pannacci2008equilibrium}. When small bubbles are injected into a liquid pool with immiscible oil droplets, bubbles with lower density rapidly rise due to buoyancy. Numerous rising bubbles touch oil droplets, and the surface tension between bubbles and droplets initiates the bubble-droplet interaction. Depending on spreading factors derived from surface tensions among three different components, three distinct bubble-droplet morphologies are expected: (1) separated bubble-droplet morphology; (2) partially engulfed morphology; and (3) completely engulfed morphology \cite{guzowski2012structure}. The stability of the bubble-droplet aggregate depends highly on the surface tension between the aggregate and the liquid pool. If the surface tension is strong enough, the aggregate will maintain its shape and continue to rise. On the contrary, a weak surface tension will induce a further deformation as the aggregate rises and the velocity of the aggregate will decrease speedily which may give rise to the break up of the aggregate.

A simple interaction of bubble and droplet occurs in a ternary flow which includes a gas bubble, an oil droplet in a liquid pool. The interaction prompts complex interface deformation and morphological change which is challenging to be tracked by any simulation methods. Among several interface tracking methods, the diffuse interface method utilizes the free energy variation which results in a thermodynamically consistent system \cite{jacqmin1996energy,baroudi2021simulation}. Featuring different free energy representations, both the Cahn Hilliard(C-H) equation and the Allen Cahn(A-C) equation \cite{cahn1959free,allen1979microscopic} have been applied to solve the phase transformation as diffuse interface methods. Compared to the A-C equation, the C-H equation keeps the mass conserved by a conservative formulation, hence, it has been widely utilized in the multi-phase flow simulation \cite{kim2005continuous,lee2010lattice,lee2005stable,abadi2018numerical}. However, the loss of mass and density shift are still observed when it is used to model a small radius bubble or droplet due to the implicit curvature-driven velocity \cite{sun2007sharp,yue2007spontaneous,zheng2014shrinkage,baroudi2021simulation}. Thus, many efforts have been made to modify the A-C equation in order to create a formulation that is both efficient and conservative. Based on the phase-field model conducted by R. Folch \cite{folch1999phase} and sharp interface tracking method investigated by Sun and Beckermann \cite{sun2007sharp}, Chiu proposed the conservative phase-field method \cite{chiu2011conservative}. The essential idea behind this method is to remain the conservation by removing the curvature-driven velocity from the A-C equation and moving the diffusion terms into the divergence operator. This modification offers a remarkable improvement in mass conservation compared to C-H equation \cite{baroudi2021simulation}. Furthermore, compared to the C-H equation where we have to solve the fourth-order partial differential equation, the conservative phase-field equation only solves a second-order partial differential equation. This feature omits the higher order derivative calculation which enables the numerical computation considerably easier. By adding a Lagrange multiplier, the model is then optimized to solve the multi-component system \cite{geier2015conservative,abadi2018conservative,aihara2019multi,zheng2020multiphase}.

According to \cite{pannacci2008equilibrium}, the equilibrium morphology and the aggregate stability mostly depend on the surface tension and the combination of different components' spreading factors. Since the surface tensions are explicitly given naturally, how to model the surface force can then be crucial during the simulation. As far as we concern, three well known surface force forms in this article are presented: (1) continuum surface force (CSF) formulation \cite{kim2005continuous,brackbill1992continuum}; (2) potential form formulation \cite{jacqmin1999calculation}; (3) stress form formulation \cite{lafaurie1994modelling}. Unlike the C-H equation, the free energy of the conservative phase-field equation is not yet complete due to the subtraction of the curvature-driven velocity. In this instance, we claim that the potential formulation which is mostly derived by energy perspective combined with the conservative phase-field equation will not be able to reduce parasitic currents as effectively as it did previously\cite{lee2006eliminating}. 
We conduct the simulations based on Lattice Boltzmann method (LBM). LBM has been widely applied to solve Navier-Stokes equation for incompressible flow \cite{lee2005stable,lee2003eulerian,lee2001characteristic} and shown as an effective method to solve the multi-phase flow problem by pseudo-potential LBM \cite{shan1993lattice} and phase-field LBM \cite{li2016lattice,lee2010lattice,abadie2015combined,abadi2018conservative}. Previously, high density and viscosity ratio are encountered when the LBM is utilized to model the multi-phase flow. To increase the system's instability, Lee introduces the multi-step collision and mixed difference method \cite{lee2010lattice}. For incompressible two-phase flow, Inamuro suggests the free energy LBM \cite{inamuro2004lattice}. Zu and He propose the velocity-based LBM to solve the high-density ratio problems \cite{zu2013phase}. It is noted that the previous model cannot entirely recover the continuity equation, and to improve this, a new velocity-pressure based LBM is proposed to deal with the problem of large ratio parameters \cite{baroudi2021simulation}. The distribution function is modified to recover the pressure evolution equation and momentum equations.

In this study, we combine the conservative phase-field LBM with the velocity-pressure based LBM. The following section goes over the derivation specifics for each LBM. In terms of the simulation, the benchmark problems including parasitic currents and liquid lens are conducted to validate the conservative character and the accuracy of the recent model. We further investigate the morphology changing problem and post the results into a diagram. Then, the dynamics of single rising bubble is investigated and the convergence test is conducted using present method. Finally we present the simulation on the rising bubble and droplet interaction. The stability and the terminal velocity of different morophologies are tested under different $Bo$.

\section{Conservative Phase-field Lattice Boltzmann Equation}

\subsection{Conservative Phase-field Equation}
The two-component conservative phase-field equation can be derived either by the free energy approach from the Allen Cahn equation~\cite{aihara2019multi} or by the velocity-based approach from the generic interface advection equation~\cite{sun2007sharp}. In the following section, the derivation based on the velocity-based approach is described~\cite{geier2015conservative}.
\subsubsection{Conservative Phase-field Equation for Two-component Flows}
Consider the following interface advection equation for a two-phase flow system:
\begin{equation}\label{Aeq:1}
    \frac{\partial\phi}{\partial t}+\boldsymbol{u}\cdot\nabla\phi=0,
\end{equation}
where the order parameter $\phi$ with the constraint $0\leq \phi \leq 1$ is used to denote different fluid's regions.     The flow velocity is represented by $\boldsymbol{u}$ which can be divided into a normal velocity $\boldsymbol{u}_n$ and an external advection velocity $\boldsymbol{u}_e$ as follow:
\begin{equation}\label{Aeq:2}
   \boldsymbol{u}=\boldsymbol{u}_n+\boldsymbol{u}_e.
\end{equation}
The normal velocity $\boldsymbol{u}_n$ can be further decomposed as:
\begin{equation}\label{Aeq:3}
   \boldsymbol{u}_n=-M\kappa\boldsymbol{n},
\end{equation}
where $M$ is the mobility which is a pure calculation parameter, $\kappa$ denotes the interface curvature and $\boldsymbol{n}$ represents the unit normal vector.  $\boldsymbol{n}$ and $\kappa$ can be expressed as a function of the order parameter $\phi$:
\begin{equation}\label{Aeq:4}
    \boldsymbol{n}=\frac{\nabla \phi}{|\nabla \phi|},
\end{equation}

\begin{equation}\label{Aeq:5}
  \kappa=\nabla\cdot\boldsymbol{n}=
\frac{1}{|\nabla\phi|}\left[\nabla^2\phi-\frac{\nabla\phi\cdot\nabla|\nabla\phi|}{|\nabla\phi|}\right].
\end{equation}

Substituting Eq.~(\ref{Aeq:4}) and Eq.~(\ref{Aeq:5}) into Eq.~(\ref{Aeq:1}), we can reformulated Eq.~(\ref{Aeq:1}) as:
\begin{equation}\label{Aeq:6}
    \frac{\partial\phi}{\partial t}+\boldsymbol{u}_e\cdot\nabla\phi=M\kappa|\nabla\phi|=M\left[\nabla^2\phi-\frac{\nabla\phi\cdot\nabla|\nabla\phi|}{|\nabla\phi|}\right]=M\left[\nabla^2\phi-\boldsymbol{n}\cdot\nabla|\nabla\phi|\right].
\end{equation}
It is noteworthy that Eq.~(\ref{Aeq:6}) is not in a conservative form and thus will induce mass conservation error. To overcome this, Folch \emph{et al.}~\cite{folch1999phase} proposed to explicitly remove the curvature driven part from Eq.~(\ref{Aeq:6}), which leads to:
\begin{equation}\label{Aeq:7}
    \frac{\partial\phi}{\partial t}+\boldsymbol{u}_e\cdot\nabla\phi=M\left[\nabla^2\phi-\boldsymbol{n}\cdot\nabla|\nabla\phi|-|\nabla\phi|\nabla\cdot\boldsymbol{n}\right]\approx M\left[\nabla^2\phi-\nabla\cdot(|\nabla\phi^{eq}|\boldsymbol{n})\right].
\end{equation}
Here $\phi^{eq}$ is the equilibrium profile of the order parameter for a planar interface, which is represented by a hyperbolic tangent function as follows:
\begin{equation}\label{Aeq:8}
  \phi^{eq}=\frac{1}{2}\left[1+\tanh\left(\frac{2z}{\delta}\right)\right],
\end{equation}
where $z$ is the normal distance between a local point and the interface with the interface thickness being adjusted by $\delta$. Eq.~(\ref{Aeq:8}) results in:
\begin{equation}\label{Aeq:9}
  |\nabla\phi^{eq}|=\frac{\partial \phi^{eq}}{\partial \bf{n}}=\frac{4\phi(1-\phi)}{\delta}.
\end{equation}
Once Eq.~(\ref{Aeq:9}) is substituted into Eq.~(\ref{Aeq:7}) and the continuity condition $\nabla \cdot \boldsymbol{u}_e =0$ is imposed, we arrive at the conservative phase-field equation for two-phase flow:
\begin{equation}\label{Aeq:10}
    \frac{\partial\phi}{\partial t}+\nabla\cdot(\phi\boldsymbol{u_e})=\nabla\cdot M\left[\nabla\phi-
    \frac{4\phi(1-\phi)/\delta}{|\nabla\phi|}\nabla\phi\right].
\end{equation}

\subsubsection{Conservative Phase Field Equation for Multi-component Flows}
The following is how we arrive at the conservative phase field equation for multi-component flows. Based on the two phase flow model, we further introduce the Lagrange multiplier $\psi_i$ \cite{kim2007phase} to the original two-component flow model to satisfy the constraint of the multi-component systems. In the following derivation, $\phi_i$ represents the order parameter of the $i^{th}$ component in the \emph{n}-component flow. We start with:
\begin{equation}\label{Aeq:11}
    \frac{\partial\phi_i}{\partial t}+\nabla\cdot(\phi_i\boldsymbol{u})=\nabla\cdot M\left(\nabla\phi_i-
    \frac{4\phi_i(1-\phi_i)/\delta}{|\nabla\phi_i|}\nabla\phi_i+\psi_i\right).
\end{equation}
In order to determine the Lagrange multiplier $\psi_i$, we first consider a system at the equilibrium, for which the left-hand side of Eq.~(\ref{Aeq:11}) disappears. The summation of the phase field equations can be calculated as:
\begin{equation}\label{Aeq:12}
  \sum_{i=1}^n\psi_i=\sum_{i=1}^{n}\frac{4\phi_i(1-\phi_i)/\delta}{|\nabla\phi_i|}\nabla\phi_i.
\end{equation}
Following the method of the derivation of this Lagrange multiplier proposed by Kim \cite{lee2015efficient}, we assume the factor before the sum calculus as:
\begin{equation}\label{Aeq:13}
  \psi_i=\frac{\phi_i^2}{\sum_{j=1}^n\phi_j^2}\sum_{j=1}^n\frac{4\phi_j(1-\phi_j)/\delta}{|\nabla\phi_j|}\nabla\phi_j.
\end{equation}

Finally, the conservative phase field equation for the multi-component flow is derived as:
\begin{equation}
 \frac{\partial\phi_i}{\partial t}+\nabla\cdot(\phi_i\boldsymbol{u})= \nabla\cdot M\left(\nabla\phi_i-\frac{4\phi_i(1-\phi_i)/\delta}{|\nabla\phi_i|}\nabla\phi_i
    +\frac{\phi_i^2}{\sum_{j=1}^n\phi_j^2}\sum_j\frac{4\phi_j(1-\phi_j)/\delta}{|\nabla\phi_j|}\nabla\phi_j\right).
 \label{Aeq:14}
\end{equation}

\subsection{\label{sec:surf}Formulations for Surface Tension Force}
The momentum equation can be expressed as follows
\begin{equation}\label{Beq:1}
    \frac{\partial \boldsymbol{u}}{\partial t}+\nabla\cdot(\boldsymbol{uu})=-\frac{1}{\rho}\nabla p +\frac{1}{\rho}\nabla\cdot \eta\left(\nabla\boldsymbol{u}+(\nabla\boldsymbol{u})^T\right)+\frac{1}{\rho}\boldsymbol{F}_s  +\frac{1}{\rho}\boldsymbol{F}_b,
\end{equation}
where $\rho$ and $\eta$ represent the density and dynamic viscosity of a mixture. In Eq.~(\ref{Beq:1}), $p$ is the dynamic pressure, $\boldsymbol{F}_s $ is the surface tension force, and $\boldsymbol{F}_b$ is the body force. We briefly list and compare the following formulations of surface tension force for two-phase flow
\begin{equation}\label{Beq:2}
    \boldsymbol{F}_{s1}=-\frac{3\sigma\delta}{2}\nabla\cdot\left(\frac{\nabla\phi}{|\nabla\phi|}\right)   |\nabla\phi|^2  \frac{\nabla\phi}{|\nabla\phi|},
\end{equation}
\begin{equation}\label{Beq:3}
    \boldsymbol{F}_{s2}=\mu\nabla\phi,
\end{equation}
\begin{equation}\label{Beq:5}
    \boldsymbol{F}_{s3}=\frac{3\sigma\delta}{2}\nabla\cdot\left(|\nabla\phi|^2I-\nabla\phi\otimes\nabla\phi\right).
\end{equation}
Here $\sigma$ represents the surface energy between two fluids. In Eq.~(\ref{Beq:3}), $\mu$ denotes the chemical potential that can be expressed as $\mu=\mu_0-\epsilon\nabla^2\phi$, and $\mu_0=\partial E_0/\partial\phi$ where $E_0=\beta\phi^2(\phi-1)^2$ is the bulk energy. $\beta$ is a constant that can be calculated from $\beta=8\epsilon/\delta ^2$ and related to the surface tension $\sigma = \sqrt{2\epsilon \beta}/6$.

Among three formulas, $\boldsymbol{F}_{s1}$ is proposed by Brackbill \cite{brackbill1992continuum} as the continuum surface force (CSF). This model, as shown in Eq. \ref{Beq:2}, calculates the curvature by an explicit derivative of the order parameter for which the performance highly depends on the derivative calculation. $\boldsymbol{F}_{s2}$ is the potential form formulation \cite{jacqmin1996energy,jacqmin1999calculation}. $\boldsymbol{F}_{s3}$, the stress form formulation, is proposed by Lafaurie \cite{lafaurie1994modelling} which is the only formulation that conserves the momentum by the divergence operator. The potential form of surface tension force are mostly applied in LBM coupled with C-H to decrease the parasitic currents by balancing the thermodynamic pressure \cite{lee2006eliminating,lee2009effects}. It is known that the phase field equations including C-H equation and A-C equation could be derived from the free energy approach. However, we should notice that, compared to the original A-C equation and the C-H equation, the conservative phase-field equation subtracts the curvature-driven term $\kappa |\nabla\phi|$ from the original free energy in A-C equation. In this case, we argue that it is not consistent to calculate the surface tension force from the chemical potential in the momentum equation.

\subsection{\label{sec:citeref}Lattice Boltzmann equations}
Through above derivations, the governing equations for the ternary flow can be expressed as the pressure evolution equation, the velocity equation and the conservative phase-field equations:
\begin{equation}\label{Ceq:1}
\frac{\partial \bar{p}}{\partial t}+\boldsymbol{u}\cdot\nabla \bar{p}+ c_s^2 \nabla\cdot \boldsymbol{u}=0,
\end{equation}
\begin{equation}\label{Ceq:2}
\frac{\partial \boldsymbol{u}}{\partial t}+\nabla\cdot(\boldsymbol{uu})=-\frac{1}{\rho}\nabla P+\frac{1}{\rho}\nabla\cdot \eta\left(\nabla\boldsymbol{u}+(\nabla\boldsymbol{u})^T\right)+\frac{1}{\rho}\boldsymbol{F}_s  +\frac{1}{\rho}\boldsymbol{F}_b,
\end{equation}
\begin{widetext}
\begin{equation}
 \frac{\partial\phi_i}{\partial t}+\nabla\cdot(\phi_i\boldsymbol{u})= \nabla\cdot M\left(\nabla\phi_i-\frac{4}{\delta}\frac{\nabla\phi_i}{|\nabla\phi_i|}\phi_i(1-\phi_i)
    +\frac{\phi_i^2}{\sum_{j=1}^3\phi_j^2}\sum_{j=1}^3\frac{4}{\delta}\frac{\nabla\phi_j}{|\nabla\phi_j|}\phi_j(1-\phi_j)\right).
\label{Ceq:3}
\end{equation}
\end{widetext}
In ternary flow, we normally solve two equations to calculate $\phi_1$, $\phi_2$ and obtain the third-order parameter $\phi_3$ from the relation equation: $\sum_{i}\phi_i=1$.

\subsubsection{Lattice Boltzmann equation for conservative phase-field equation}
The comprehensive derivation of LBM for the conservative phase-field equation is offered in this section. The Discrete Boltzmann equation (DBE) for ternary flow can be represented as:
\begin{equation}\label{Ceq:4}
\left(\frac{\partial}{\partial t}+\boldsymbol{e}_\alpha\cdot\nabla\right)h_\alpha^i=
   -\frac{1}{\lambda_\phi}(h_\alpha^i-h_\alpha^{i,eq})+\Gamma_\alpha(\boldsymbol{e}_\alpha-\boldsymbol{u})\cdot\boldsymbol{S_i},
\end{equation}
where $i=1,2,3$, $h^i_\alpha$ and $h_\alpha^{i,eq}$ represent the particle distribution function and equilibrium distribution function for $i^{th}$ component order parameter. $\boldsymbol{e}_\alpha$ denotes the lattice velocity in $D2Q9$ lattice given as:
$$ \boldsymbol{e}_\alpha=\left\{
\begin{aligned}
&(0,0)c,&     &\alpha=0 \\
&(cos\theta_\alpha,sin\theta_\alpha)c,& \theta_\alpha=(\alpha-1)\pi/2, \quad &\alpha=1,2,3,4\\
&\sqrt{2}(cos\theta_\alpha,sin\theta_\alpha)c,& \theta_\alpha=(\alpha-5)\pi/2+\pi/4, \quad &\alpha=5,6,7,8
\end{aligned}
\right.
$$
where $c$ represents the lattice velocity unit. $\lambda_\phi$ is the relaxation time relevant to the mobility $M=\lambda_\phi c_s^2$, $c_s=\frac{1}{\sqrt{3}}c$ is the speed of sound. The equilibrium distribution function $h_\alpha^{i,eq}$ takes the form:
\begin{equation}\label{Ceq:5}
   h_\alpha^{i,eq}=t_\alpha\phi_i\left[1+\left(\frac{\boldsymbol{e}_\alpha\cdot\boldsymbol{u}}{c_s^2}+\frac{\left(\boldsymbol{e}_\alpha\cdot\boldsymbol{u}\right)^2}{2c_s^4}-\frac{\boldsymbol{u}\cdot\boldsymbol{u}}{2c_s^2}\right)\right].
\end{equation}
$t_\alpha$ is the weight with the value: $t_0=4/9, t_1=t_3=t_5=t_7=1/9$ and $t_2=t_4=t_6=t_8=1/36$. $\Gamma_\alpha$ can be calculated as $\Gamma_\alpha=h_\alpha^{i,eq}/\phi_i$. $\boldsymbol{S_i}$ is the source term from the governing equation. The macroscopic equation recovered by Chapman-Enskog expansion is then:
\begin{equation}\label{Ceq:6}
    \frac{\partial\phi_i}{\partial t}+\nabla\cdot(\phi_i\boldsymbol{u})= \nabla\cdot M\left(\nabla\phi_i-\frac{4}{\delta}\frac{\nabla\phi_i}{|\nabla\phi_i|}\phi_i(1-\phi_i)
    +\frac{\phi_i^2}{\sum_{j=1}^3\phi_j^2}\sum_{j=1}^3\frac{4}{\delta}\frac{\nabla\phi_j}{|\nabla\phi_j|}\phi_j(1-\phi_j)\right).
\end{equation}
where the source term $\boldsymbol{S}_i$ for component $i$ can be expressed as:
\begin{equation}\label{Ceq:7}
\boldsymbol{S}_i=\frac{4 }{\delta}\frac{\nabla\phi_i}{|\nabla\phi_i|}\phi_i(1-\phi_i)-\frac{\phi_i^2}{\sum_{j=1}^3\phi_j^2}\sum_{j=1}^3\frac{4 }{\delta}\frac{\nabla\phi_j}{|\nabla\phi_j|}\phi_j(1-\phi_j).
\end{equation}
The recovered phase field equation is identical with the proposed phase field equation Eq.\ref{Ceq:3}. Then, we start to solve Eq.\ref{Ceq:4} for i component by the time integration in $[t,t+\delta t]$:
\begin{multline}\label{Ceq:8}
h^i_\alpha(\boldsymbol{x}+\delta t \boldsymbol{e}_\alpha,t+\delta t)-h^i_\alpha(\boldsymbol{x},t)=
    -\int_t^{t+\delta t}\frac{h^i_\alpha-h^{i,eq}_\alpha}{\lambda_\phi}dt+\\
    \int_t^{t+\delta t} \Gamma_\alpha(\boldsymbol{u})(\boldsymbol{e}_\alpha-\boldsymbol{u})\cdot\boldsymbol{S}_i dt.
\end{multline}
Using trapezoidal rule, the time discretized equation becomes:
\begin{multline}\label{Ceq:9}
    h^i_\alpha(\boldsymbol{x}+\delta t \boldsymbol{e}_\alpha,t+\delta t)-h^i_\alpha(\boldsymbol{x},t)=
    -\frac{h^i_\alpha-h^{i,eq}_\alpha}{2\tau_\phi}\big|_t
    -\frac{h^i_\alpha-h^{i,eq}_\alpha}{2\tau_\phi}\big|_{t+\delta t}\\
    +\frac{\delta t}{2}\left[\Gamma_\alpha(\boldsymbol{u}(\boldsymbol{e}_\alpha-\boldsymbol{u})\cdot\boldsymbol{S}_i )\big|_t+
    \Gamma_\alpha(\boldsymbol{u})(\boldsymbol{e}_\alpha-\boldsymbol{u})\cdot\boldsymbol{S}_i\big|_{t+\delta t}\right].
\end{multline}
Here $\tau_\phi=\lambda_\phi/\delta t$ is dimensionless relaxation time.
We introduce the modified distribution function $\bar{h}^i(\boldsymbol{x},t)$ :
\begin{equation}\label{Ceq:10}
\bar{h}^i_\alpha(\boldsymbol{x},t)=h^i_\alpha(\boldsymbol{x},t)+\frac{1}{2\tau_\phi}\left(h^i_\alpha-h_\alpha^{i,eq}\right)\bigg|_t-\frac{\delta t}{2} \Gamma_\alpha(\boldsymbol{u}) (\boldsymbol{e}_\alpha-\boldsymbol{u})\cdot\boldsymbol{S}_i\bigg|_t.
\end{equation}
The LBM with modified distribution function for phase field equation can be written as:
\begin{multline}\label{Ceq:11}
     \bar{h}^i_\alpha(\boldsymbol{x}+\boldsymbol{e}_\alpha\delta t,t+\delta t)-\bar{h}^i_\alpha(\boldsymbol{x},t)=
     -\frac{1}{\tau_\phi+0.5}(\bar{h}^i_\alpha(\boldsymbol{x},t)-\bar{h}_\alpha^{i,eq})\\
     +\delta t \Gamma_\alpha(\boldsymbol{u})(\boldsymbol{e}_\alpha-\boldsymbol{u})\cdot\boldsymbol{S}_i.
\end{multline}
The equilibrium modified distribution function can be calculated by:

\begin{equation}\label{Ceq:12}
   \bar{h}^{i,eq}_\alpha = h^{i,eq}_\alpha
-\frac{\delta t}{2}\Gamma_\alpha(\boldsymbol{u}) (\boldsymbol{e}_\alpha-\boldsymbol{u})\cdot\boldsymbol{S}_i.
\end{equation}
\subsubsection{velocity-pressure based Lattice Boltzmann equation}
The velocity-based Lattice Boltzmann equation for high density and viscosity contrasts is proposed in \cite{zu2013phase}. It is then applied to the conservative phase-field method \cite{fakhari2017improved,abu2018conservative}. In our simulation, we use a velocity-pressure-based LBM, in which the distribution function is modified to recover the pressure \cite{baroudi2021simulation}.

The DBE for velocity-pressure formulation is given as:
\begin{equation}\label{Ceq:13}
    \left(\frac{\partial }{\partial t}+\boldsymbol{e}_\alpha \cdot\nabla\right)g_\alpha =-\frac{1}{\lambda}(g_\alpha-g_\alpha^{eq})+F_\alpha.
\end{equation}
The Chapman-Enskog expansion based on this DBE is given in Appendix A. The governing equations, Eq.\ref{Ceq:1} and Eq.\ref{Ceq:2}, can be recovered from the DBE Eq.\ref{Ceq:13}.
Following the same procedure of phase-field LBM derivation, The velocity-pressure based LBM is then given as:
\begin{equation}\label{Ceq:14}
    \bar{g}_\alpha(\boldsymbol{x}+\boldsymbol{e}_\alpha\delta t,t+\delta t)-\bar{g}_\alpha(\boldsymbol{x},t)=-\frac{1}{\tau_\rho+0.5} (\bar{g}_\alpha(\boldsymbol{x},t)-\bar{g}^{eq}_\alpha)+\delta t F_\alpha,
\end{equation}
where $\tau_\rho$ is the dimensionless relaxation time, and $\bar{g}_\alpha^{eq} $ is the modified distribution function:
\begin{equation}\label{Ceq:15}
\bar{g}_\alpha^{eq}=g_\alpha^{eq}-\frac{1}{2}F_\alpha,
\end{equation}
\begin{equation}\label{Ceq:16}
g_\alpha^{eq}=t_\alpha\bar{p}+\Gamma_\alpha c_s^2 -t_\alpha c_s^2,
\end{equation}
where $P$ represents the dynamic pressure and $\rho$ represents the local density. $\bar{p}$ can be calculated as $\bar{p}=\frac{P}{\rho}$.  The source term of Eq.\ref{Ceq:12} is composed by a collection of forcing terms:
\begin{multline}\label{Ceq:17}
F_\alpha=-\Gamma_\alpha (\boldsymbol{e}_\alpha-\boldsymbol{u})\cdot\left(\frac{1}{\rho}\nabla P\right)+\Gamma(0)(\boldsymbol{e}_\alpha-\boldsymbol{u})\cdot\left(\nabla \bar{p}\right)+\\\Gamma_\alpha (\boldsymbol{e}_\alpha-\boldsymbol{u})\cdot\left[\frac{\nu}{\rho}(\nabla\boldsymbol{u}+\nabla\boldsymbol{u}^T)\nabla\rho+\frac{1}{\rho}\boldsymbol{F}_s+\frac{1}{\rho}\boldsymbol{F}_b\right].
\end{multline}
$\boldsymbol{F}_s$ and $\boldsymbol{F}_b$ represent the surface tension force and the body force. The CSF formulation, Eq.\ref{Beq:3}, is applied in our approach. Here we only consider the gravitational force as the body force which:
\begin{equation}\label{Ceq:18}
\boldsymbol{F}_b=(\rho-\rho_{l})\boldsymbol{g},
\end{equation}
where $\boldsymbol{g}$ is the gravitation acceleration. $\rho$, $\rho_{l}$ represent the local fluid density and the background liquid density.
The derivatives of macroscopic value which appear in Eq.\ref{Ceq:17} can be calculated by second order isotropic finite difference method \cite{lee2005stable}:
\begin{equation}\label{Ceq:19}
  \frac{\partial \phi}{\partial x_i}=\sum_{\alpha\neq0}\frac{t_\alpha \boldsymbol{e}_\alpha \cdot \hat{\boldsymbol{i}}[\phi(\boldsymbol{x}+\boldsymbol{e}_\alpha \delta t)-\phi(\boldsymbol{x}-\boldsymbol{e}_\alpha \delta t)] }{2c_s^2\delta t}.
\end{equation}
Then $\nabla \rho$ can be calculated from $\nabla \phi$ which:
\begin{equation}\label{Ceq:20}
\nabla \rho=\sum_{i=1}^{3}\rho_i\nabla\phi_i.
\end{equation}
After the collision and the propagation, we need to update the macroscopic value and parameters from the distribution function.

We first update different order parameters:
\begin{equation}\label{Ceq:21}
\phi_i=\sum_{\alpha=0}^8 \bar{h}_\alpha^i.
\end{equation}
The density is then updated as:
\begin{equation}\label{Ceq:22}
\rho=\sum_{i=1}^{3}\rho_i\phi_i.
\end{equation}
After that, the local viscosity and the relaxation time for distribution function $g_\alpha$ can be updated by:
\begin{equation}\label{Ceq:23}
\nu=\sum_{i=1}^{3}\nu_i\phi_i,
\end{equation}
\begin{equation}\label{Ceq:24}
\tau_\rho=\frac{\nu}{ c_s^2\delta t}.
\end{equation}
In the end, the macroscopic value of pressure and the velocity can be calculated from the zero and first moment of the distribution function $g_\alpha$:
\begin{equation}\label{Ceq:25}
\bar{p}=\sum_{\alpha}\bar{g}_\alpha+\frac{\delta t}{2}\sum_\alpha F_\alpha,
\end{equation}
\begin{equation}\label{Ceq:26}
\boldsymbol{u}=\sum_{\alpha}\frac{\bar{g}_\alpha}{c_s^2}  \boldsymbol{e}_\alpha +\frac{\delta t}{2c_s^2}\sum_\alpha F_\alpha \boldsymbol{e}_\alpha.
\end{equation}

\section{Numerical tests}
The primary parameters appear in simulations are the diameter of a bubble or a droplet $D$, the dynamic viscosity of the $i^{th}$ component $\eta_i$, and the surface tension between $i^{th}$ and $j^{th}$ components $\sigma_{ij}$. They are used to calculate the dimensionless groups, which are summarized as follows: 

$$Cn=\frac{\delta}{D},$$
$$La=\frac{\sigma_{ij}\rho D}{\eta_{i}^2},$$
$$Bo=\frac{\Delta\rho_i g D^2}{\sigma_{ij}},$$
$$Ar=\frac{\rho_i\sqrt{gD^3}}{\eta_i}, $$
$$Oh=\frac{\eta_{i}}{\sqrt{\rho_i\sigma_{ij} D}},$$
where $\delta$ is the interface thickness between two fluids, $Cn$, the Cahn number, which is defined as the ratio of the unphysical interface thickness and the diameter. It is mostly used to evaluate if the phase field method achieves a sharp interface limit by convergence test. $La$ denotes the Laplace number which estimate the surface tension and momentum effect. $Bo$ and $Ar$ are referred to as the Bond number (also known as E\"{o}tv\"{o}s number) and the Archimedes number. These two parameters are introduced to monitor the dynamics of a rising bubble under the gravity field \cite{liang2019axisymmetric,hua2007numerical}. The Ohnesorge number, $Oh$, is a measure of the strength of the interaction between the bubble and droplet. For inertia regime, we have $Oh\ll1$, when the fluids are brought to contact, there will exist a significant fluid-fluid interaction at the interface region. When $Oh\gg1$, the interaction will be smoothed by fluids' viscosity. The spreading factor for the $k^{th}$ component can be calculated from the surface tensions, for example $S_k=\sigma_{ij}-(\sigma_{ik}+\sigma_{jk})$. Partially engulfed, double emulsion and separate terminal morphology can be expected when we have relative combination of spreading factors.

\subsection{Parasitic Currents}

The first test case we considered is a stationary droplet inside a liquid pool. A stationary droplet is placed in the middle of a quiescent fluid without gravity field, and ideally the velocity magnitude is expected to remain zero. However, due to the numerical error and the unbalance pressure occurs when the surface force coming to the system, the unphysical parasitic currents appear at the interface region in simulation \cite{popinet2018numerical}. As a result, how to apply this surface force is critical to simulate multi-phase flow. For this test, we aim to distinguish different types of surface tension formulations which have been proposed previously and to examine their performance under different system parameters. Initially, the droplet with $D/\Delta x=50 $ is centered in a square domain $L/D=2$ with a fixed density and viscosity ratio $\rho*=\rho_1/\rho_2=1$, $\eta^*=\eta_1/\eta_2=1$. Four separate boundaries are subjected to the symmetric boundary condition. The relaxation times for momentum equation and phase field equation are set as constant: $\tau_\rho=0.5$ and $\tau_\phi=0.3$, and the surface tension between two fluids is given as $\sigma_{12}=1\times10^{-4} $ in lattice unit. Convergence test is carried out for different $Cn$ for which we expect a lower intensity of parasitic currents when we gradually increase $Cn$. To make the system achieves an equilibrium state, the simulation results are reported after $T/t_0=200$, where $t_0=\eta_1 D/2\sigma_{12}$ denotes the viscous time scale.
\begin{figure*}[hbt!]

  \centering

  \includegraphics[width=\textwidth]{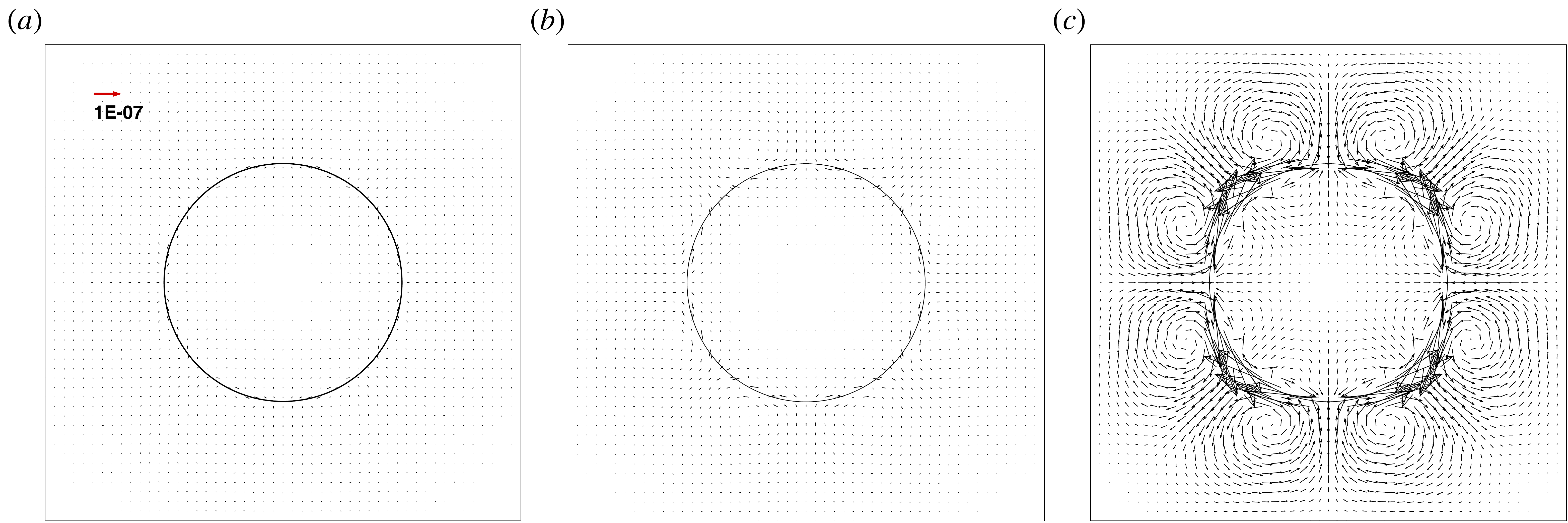}

  \caption{\label{exfig:1} The parasitic currents' vectors of the single droplet simulation at $T=200t_0$ by using (a) the continuous surface formulation; (b) the potential formulation; (c) the stress formulation, for $Cn=0.08$. The interface is represented by a contour level $\phi=0.5$ . The reference vector with magnitude $1e-7$ is indicated by a red arrow in (a).}
\end{figure*}

To group different cases, the interface thicknesses are set as $\delta/\Delta x=[2, 3,4]$, correspondingly $Cn=[0.04, 0.06,0.08]$. Figure~\ref{exfig:1} shows the flow field velocity vector when applying different formulations with $Cn=0.08$. We provide the intensity of parasitic currents defined as $\boldsymbol{u}_{max}^2$ in Table~\ref{Tab:1}. According to Table~\ref{Tab:1}, the CSF formulation performs much better results than the potential form formulation when we have a smaller interface thickness. All of these three formulations are able to reduce the parasitic currents intensity as $Cn$ increases.

\begin{table}
\caption{\label{Tab:1}Convergence test of the parasitic currents intensity with different $Cn$.}
\begin{ruledtabular}
\begin{tabular}{lllll}
   &&
  \multicolumn{2}{l}{$\boldsymbol{u}_{max}^2$}&\\
 \cline{3-5}
  $\delta$&Cn &CSF& Potential form & Stress form \\ \hline

    2  & 0.04 &$2.6\times10^{-14}$&$2.0\times10^{-11}$ & $5.9\times10^{-13}$\\

    3  & 0.06  &$2.3\times10^{-15}$&$4.8\times10^{-14}$ &$1.4\times10^{-13}$ \\

    4  & 0.08  &$4.4\times10^{-16}$&$1.6\times10^{-15}$ &$3.2\times10^{-14}$ \\

\end{tabular}
\end{ruledtabular}

\end{table}

\begin{table}
\caption{\label{Tab:2}Convergence test of the parasitic currents intensity with  different $La$.}
\begin{ruledtabular}
\begin{tabular}{lllll}
   &
  \multicolumn{2}{l}{$\boldsymbol{u}_{max}^2$}&\\
 \cline{2-4}
  $La$ &CSF& Potential form & Stress form \\ \hline

  16    & $8.5\times10^{-14}$ &$1.1\times10^{-12}$ & $3.8\times10^{-12}$\\

  4     & $2.2\times10^{-14}$ &$3.2\times10^{-13}$ & $1.1\times10^{-12}$\\

  1     & $5.3\times10^{-15}$ &$1.1\times10^{-13}$ &$4.3\times10^{-13}$\\

  0.25  & $2.4\times10^{-15}$ &$5.1\times10^{-14}$ &$1.8\times10^{-13}$\\

\end{tabular}
\end{ruledtabular}
\end{table}

\begin{table}
\caption{\label{Tab:3}Convergence test of the parasitic currents intensity based on coupled momentum equation and conservative phase-field equation with  different $La$.}
\begin{ruledtabular}
\begin{tabular}{llll}
   &
  \multicolumn{2}{l}{$\boldsymbol{u}_{max}^2$}&\\
 \cline{2-4}
  $La$ &CSF& Potential form & Stress form \\ \hline
  16    & $5.0\times10^{-12}$ &$8.0\times10^{-12}$ & $1.2\times10^{-10}$\\

  4     & $4.5\times10^{-12}$ &$5.4\times10^{-12}$ & $3.1\times10^{-11}$\\

  1     & $3.9\times10^{-12}$ &$4.1\times10^{-12}$ & $8.7\times10^{-12}$\\

  0.25  & $2.9\times10^{-12}$ &$2.7\times10^{-12}$ & $3.5\times10^{-12}$\\

\end{tabular}
\end{ruledtabular}
\end{table}

We then simulate the parasitic currents with fixed $Cn=0.06$ and $La=[0.25-16]$. The results are posted in Table~\ref{Tab:2}. According to the simulation results, the parasitic currents intensity for all three different formulations have a decreasing trend when $La$ is decreasing. When $La=16$, which means the surface tension behaves more than momentum transport, the CSF gains 100 times smaller parasitic currents. The potential form obtains a relative quick convergence trend than both the CSF and the stress form formulation.

When coupled with the conservative phase-field equation, the effect of combination of momentum and phase field equation is presented. As well, we keep a fixed $Cn=0.06$ and conduct the convergence test with changing $La=[0.25-16]$. Here,the density ratio and viscosity ratio are introduced to the system: $\rho^*=\rho_l/\rho_g=10$, $\eta^*=\eta_l/\eta_g=10$. According to Table~\ref{Tab:3}, the CSF and potential form formulations perform better than the stress form formulation when we consider a larger $La$. If the $La$ is small enough, three formulations obtain similar results.

Through those results for only solving momentum equation, we learn that if the curvature of the droplet keeps fixed and the system is under a large surface tension effect, the CSF behaves much better than the other approaches. The reason can be explained that the CSF is the only formulation that explicitly calculate the curvature term\cite{kim2005continuous}. When the curvature keeps fixed, the performance of the CSF is greatly increased. However, when the order parameter evolves, the curvature will change over time. the CSF loses this advantage, and the performance decreases. Potential form formulation tries to balance the pressure gradient. Due to the inconsistent energy comes from the conservative phase field method, the parasitic currents still appear in the simulation.

\begin{figure}[htb!]
  \centering
  \includegraphics[width=0.5\textwidth]{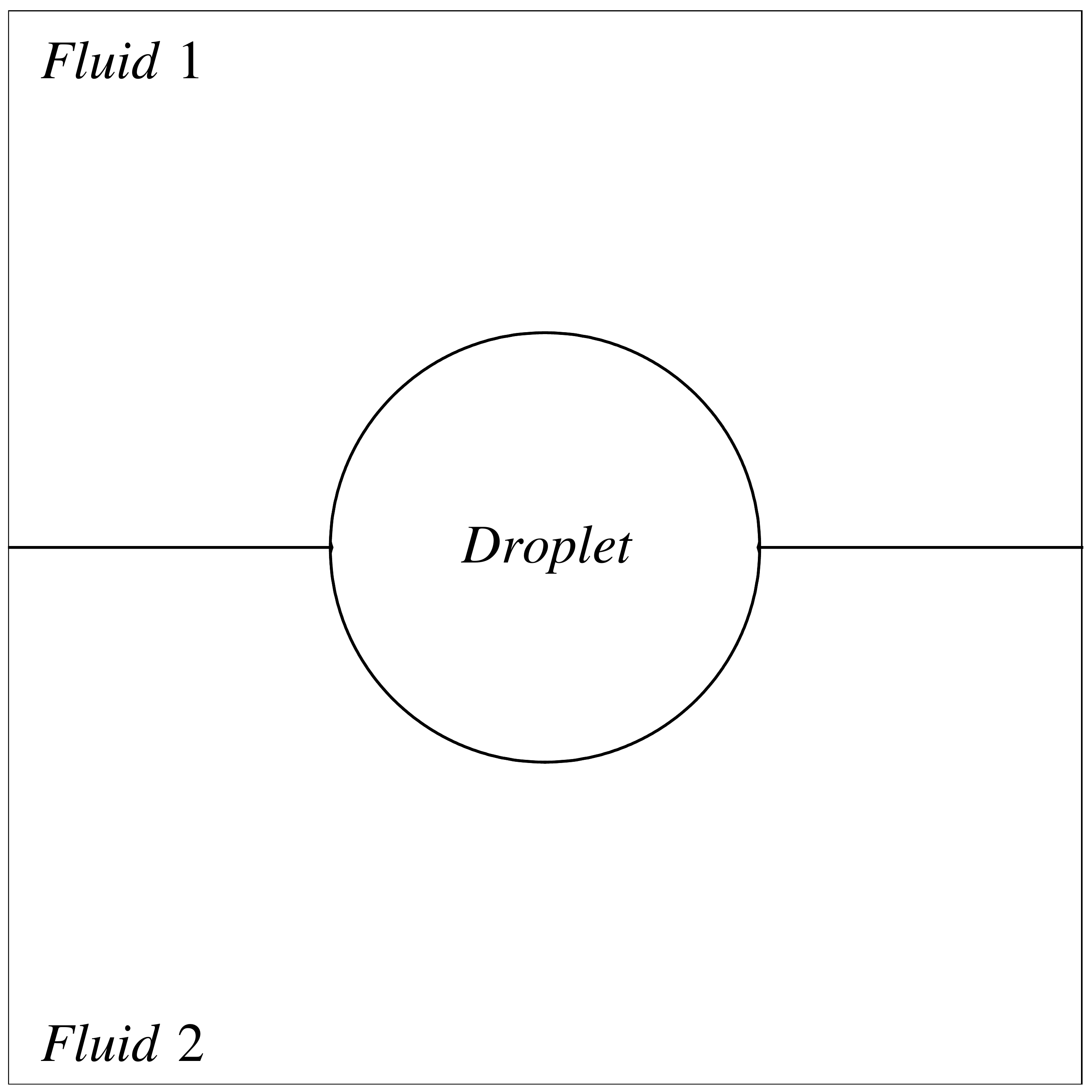}
    \caption{\label{exfig:2}Initial profile of the liquid lens simulation. A circle droplet is placed in between of two other fluids. The interface between different fluids are shown by $\phi=0.5$.}

\end{figure}

\subsection{Liquid Lens}
The liquid lens problem is widely applied as a validation case for ternary flow simulation and we present this test to show our model's capability to deal with ternary flows. The initial state is shown as Fig.~\ref{exfig:2}, where the droplet is placed into two fluids. The center of the droplet is settled in the middle of the square domain. This droplet keeps deforming due to the surface force and resisted by the viscous dissipation until arriving at the equilibrium state. By controlling the surface tension ratios, we could achieve different contact angles at triple contact points when the system arrives at equilibrium. The initial order parameter profiles are set as the functions \cite{abu2018conservative}:
\begin{equation}\label{ex:1}
\begin{split}
   \phi_2(\boldsymbol{x},0) & =\frac{1}{2}+\left[tanh\left(\frac{2}{\delta}min(|\boldsymbol{x}-\boldsymbol{x}_c|-R,y-y_c)\right)\right] ,\\
     \phi_3(\boldsymbol{x},0)&=\frac{1}{2}-\left[tanh\left(-\frac{2}{\delta}min(|\boldsymbol{x}-\boldsymbol{x}_c|+R,y-y_c)\right)\right],\\
    \phi_1(\boldsymbol{x},0) & =1-\phi_1(\boldsymbol{x},0)-\phi_2(\boldsymbol{x},0),
\end{split}
\end{equation}
where $\boldsymbol{x}_c$ is the center of the liquid lens. The basic theory of liquid lens can be expressed as Neumann's triangle. According to Neumann's theory, when the whole system reaches equilibrium state, the relation between contact angles $\theta_i$, $\theta_j$, $\theta_k$ and surface tensions of three phases $\sigma_{ij}$, $\sigma_{ik}$, $\sigma_{jk}$ are given as:
\begin{equation}\label{ex:2}
  \frac{sin\theta_i}{\sigma_{jk}}=\frac{sin\theta_j}{\sigma_{ik}}=\frac{sin\theta_k}{\sigma_{ij}},
\end{equation}
The analytic contact angle can be calculated by:
\begin{equation}
  \theta_i=cos^{-1}\left(-\frac{\sigma_{ij}^2 +\sigma_{ik}^2 +\sigma_{jk}^2}{2\sigma_{ij}\sigma_{ik}}\right).
\end{equation}\label{ex:3}

\begin{figure}[htb!]
  
  \centering
  \includegraphics[width=0.5\textwidth]{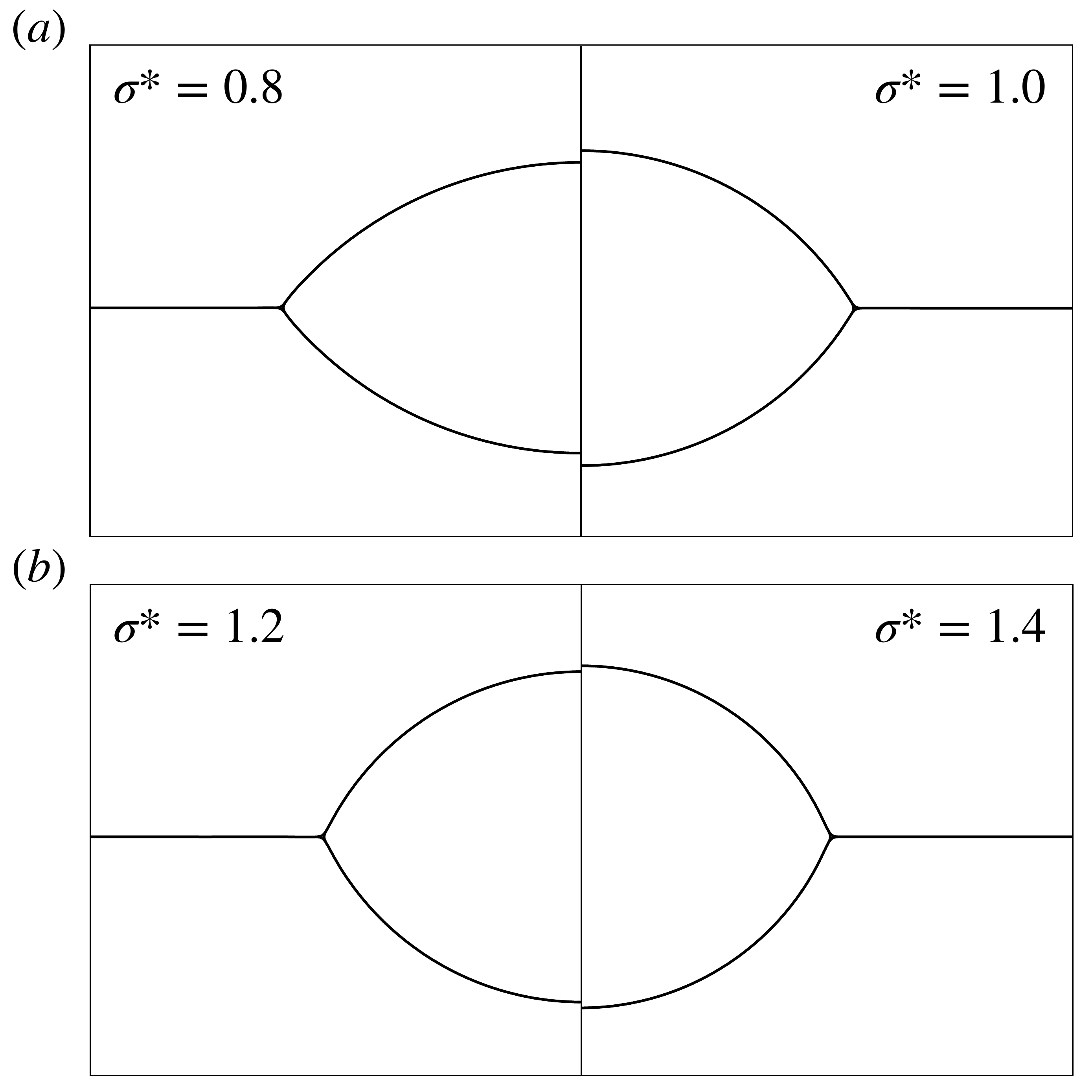}

    \caption{\label{exfig:4} Equilibrium morphology of the liquid lens simulations for various surface tensions between droplet and fluids with $Cn=0.01875$: (a) (left panel) $\sigma^*=0.8$; (right panel) $\sigma^*=1.0$; (b) (left panel) $\sigma^*=1.2$; (right panel) $\sigma^*=1.4$.  }

\end{figure}

The method of the contact angle calculation from the simulation can be found in Appendix B, by which we can compare our results with analytic solutions. We start with the given surface tension ratios $(\sigma_{12}, \sigma_{13}, \sigma_{23})\times 10^3=(1,1,0.8),$ $(1,1,1),(1,1,1.2),(1,1,1.4)$ and the equilibrium profiles of the liquid lens are shown in Fig.~\ref{exfig:4}, where $\sigma^*=\sigma_{13}/\sigma_{12}$.

\begin{figure}[htb!]

  \centering
  \includegraphics[width=0.5\textwidth]{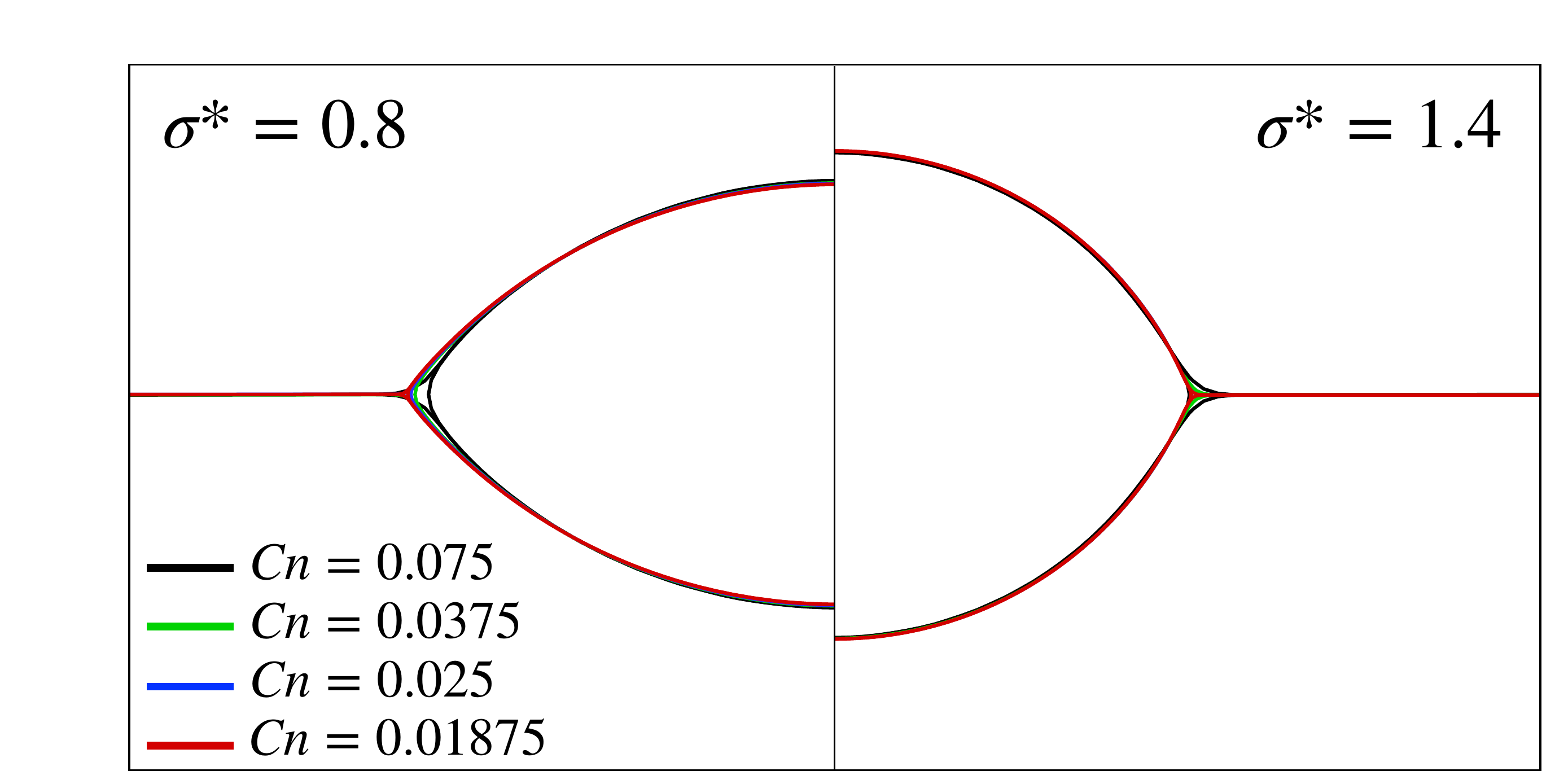}

    \caption{\label{exfig:5} Convergence test for contact lines with (left panel) $\sigma^*=0.8$ and (right panel) $\sigma^*=1.4$, $Cn=[0.01875-0.075]$. The contours of $\phi_i=0.5$ for same $Cn$ are indicated in a same color.}

\end{figure}
\begin{table*}
\caption{\label{Tab:4}The analytic solutions versus the simulation results for liquid lens length.}
\begin{ruledtabular}
\begin{tabular}{llllll}

  D/$\Delta x$   &    $40$       & $80$     & $120$      &  $160$ & analytic solution\\ \hline

$\sigma^*=0.8$  & 0.57493  &0.59372   &0.59954   & 0.60387 &0.6128\\

$\sigma^*=1.0$  & 0.53994  &0.54804   &0.55047   & 0.55192 &0.5540\\

$\sigma^*=1.2$  & 0.51734  &0.52111   &0.52185   & 0.52149 &0.5220\\

$\sigma^*=1.4$  & 0.50192  &0.50289   &0.50273   & 0.50220 &0.5014

\end{tabular}
\end{ruledtabular}

\end{table*}

The liquid lens length $d$ which is the distance between two triple contact points for equilibrium state is used to evaluate the accuracy of simulation methods. The analytic solution of $d$ is:
\begin{widetext}
\begin{equation}
\frac{1}{d^2} =\frac{1}{8A}\left(\frac{2(\pi-\theta_1)-sin(2(\pi-\theta_1)))}{sin^2(\pi-\theta_1)}
+\frac{2(\pi-\theta_3)-sin(2(\pi-\theta_3)))}{sin^2(\pi-\theta_3)}\right),
\end{equation}
\end{widetext}
where $A=\pi R^2$ is the area of initial droplet. Table~\ref{Tab:4} shows the ratio of the liquid lens length to the domain length $d/L$ with different diameters. We consider the viscosity and the density ratios ratio as: $\eta_1/\eta_2=\eta_1/\eta_3=1$, $\rho_1/\rho_2=\rho_1/\rho_3=1$. The relaxation times are set as: $\tau_\rho=0.5, \tau_\phi=3$.

Fig.~\ref{exfig:5} presents the convergence result for $\sigma^*=0.8$ and $\sigma^*=1.4$. Through the detailed results provided in Table~\ref{Tab:4}, when we decrease $Cn$, the liquid lens simulation results converge to a better value towards the analytic solution.

\subsection{Droplet Morphology}
\begin{figure}[htb!]
  \centering
  \includegraphics[width=0.4\textwidth]{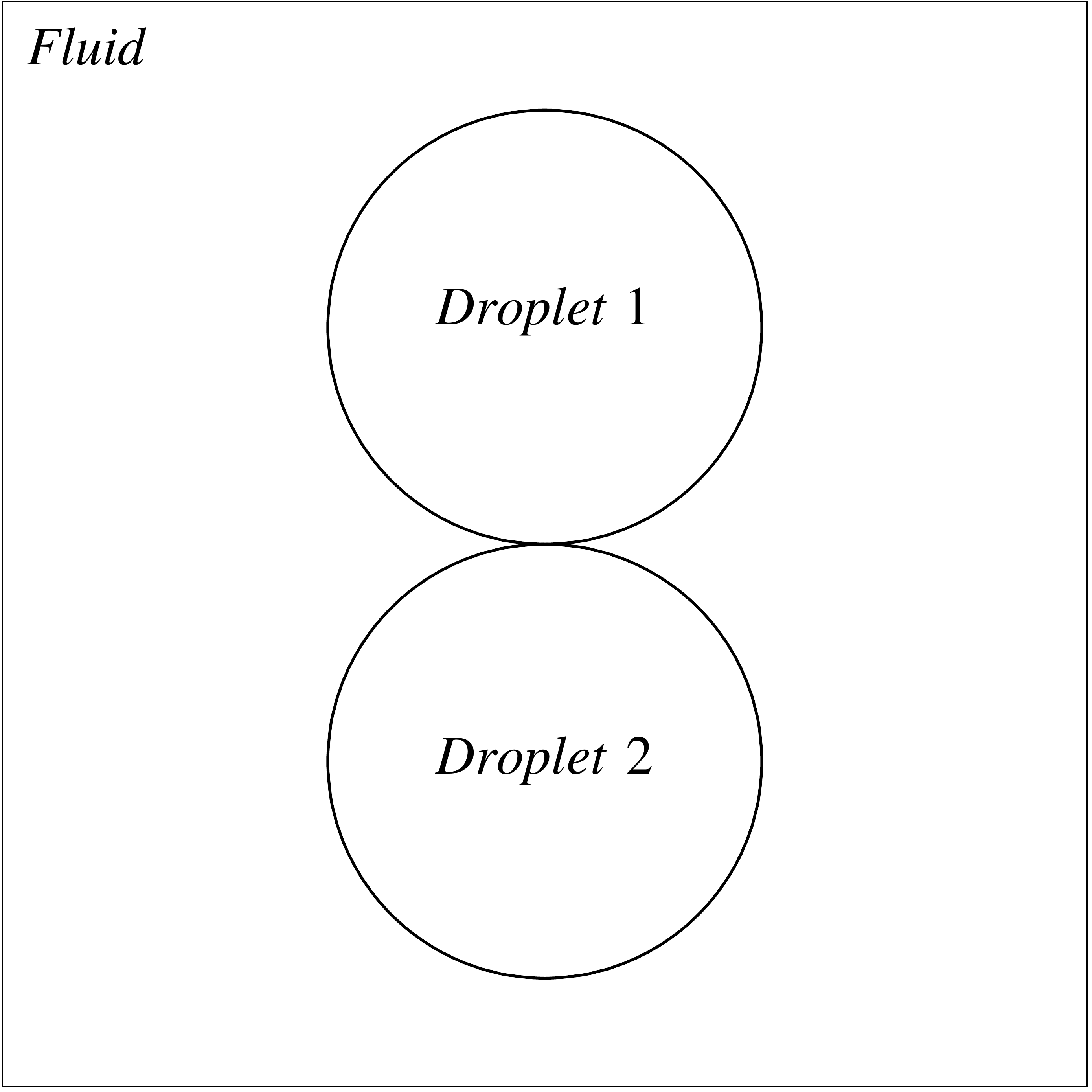}
  \caption{\label{exfig:6}Setup of the double emulsion simulation. Two equal sized droplets are initially placed inside of the background fluid. The center distance between two droplets is $C_d=D+\delta$. }
\end{figure}
\begin{figure*}[hbt!]
  \centering
  \hspace*{-1cm}
  \includegraphics[width=\textwidth]{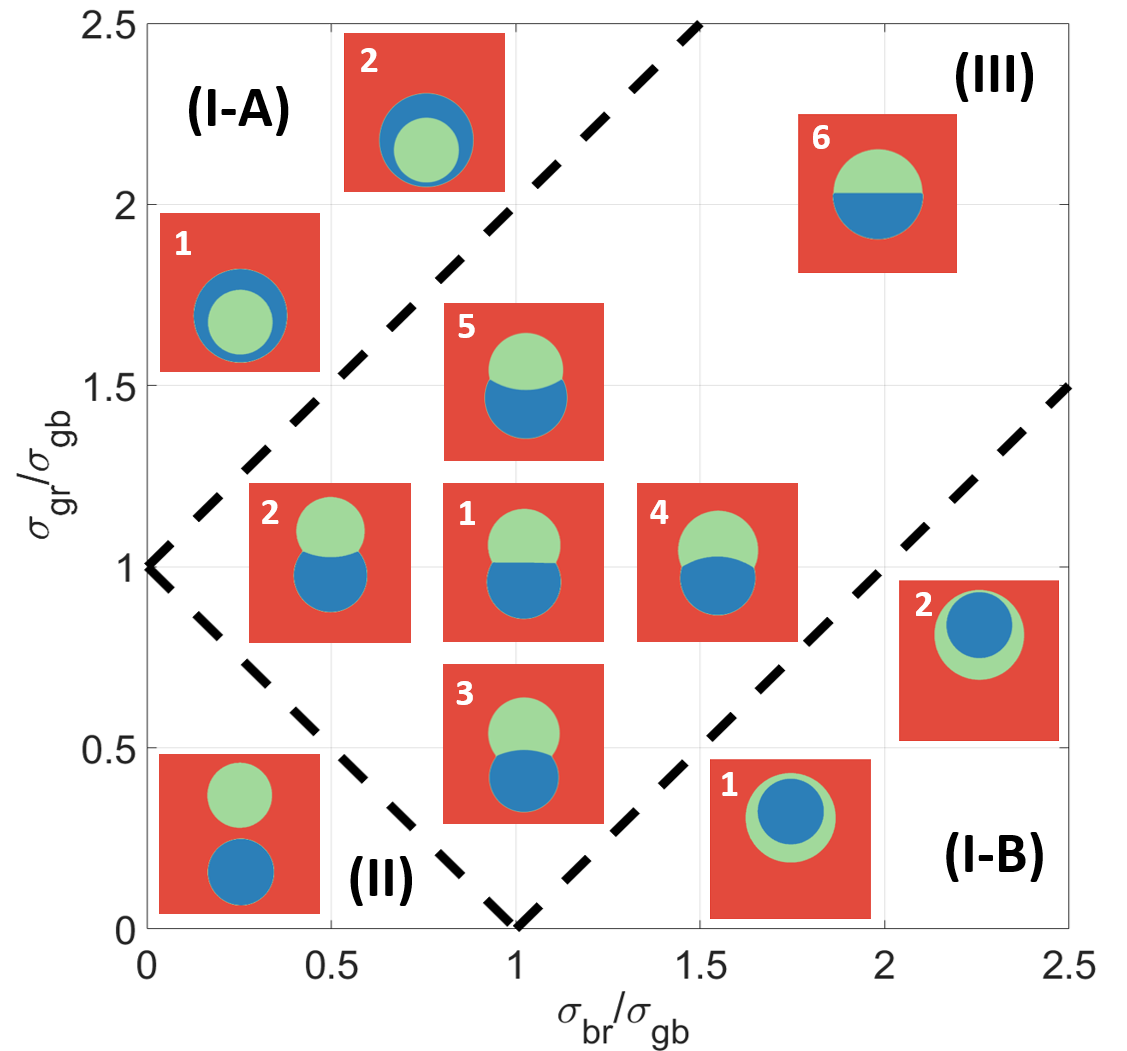}
  \caption{\label{exfig:7}Morphology diagram of the terminal shapes for double emulsion simulation. (I-A), (I-B) present the double emulsion region, (II) shows the separate morphology region, (III) represents the partially engulfed region.}
\end{figure*}
In this simulation, as Fig.~\ref{exfig:6}, we place two equal-sized droplets where $D/\Delta x=40$ into square domain $L/D=2.5$. Due to the surface force between those two droplets, we arrive at the results of different final morphology. When defining various fluid spreading phenomena, we use the spreading factor $S$. According to Pannacci \cite{pannacci2008equilibrium}, for late time morphology of two contact droplets, the complete engulfing (double emulsion) morphology appears while $S_b>0,S_g<0,S_r<0$ or $S_g>0,S_b<0,S_r<0$ corresponding to (I-A) region and (I-B) region. In (II) region, $S_r>0,S_b<0,S_g<0$, the two droplets will break up into two parts. In (III) region, $S_r<0,S_b<0,S_g<0$, the two droplets are partially engulfed by each other. Especially when we have a large surface tension ratio, the Janus droplets will appear.
\begin{table}
\caption{\label{Tab:5}Parameters diagram of double emulsion simulations. Region (I-A) and region (I-B) indicate complete engulfment morphology, Region (II) represents separated morphology and region (III) represents partially engulfed morphology.}
\begin{ruledtabular}
\begin{tabular}{cccccccc}
 Case Number&$\sigma_{gb}$&$\sigma_{br}/\sigma_{gb}$&$\sigma_{gr}/\sigma_{gb}$&$S_b$&$S_g$&$S_r$\\ \hline
 I-A(1)  & 0.05  & 0.5  & 1.55 &$>0$ &$<0$ &$<0$\\

 I-A(2)  & 0.05  & 1    & 2.05 &$>0$ &$<0$ &$<0$\\

 I-B(1)  & 0.05  & 1.55 & 0.5  &$<0$ &$>0$ &$<0$\\

 I-B(2)  & 0.05  & 2.05 & 1    &$<0$ &$>0$ &$<0$\\

 II      & 0.05  & 0.35 & 0.35 &$<0$ &$<0$ &$>0$\\

 III(1)  & 0.05  & 1    & 1    &$<0$ &$<0$ &$<0$\\

 III(2)  & 0.05  & 0.5  & 1    &$<0$ &$<0$ &$<0$\\

 III(3)  & 0.05  & 1    & 0.5  &$<0$ &$<0$ &$<0$\\

 III(4)  & 0.05  & 1    & 1.5  &$<0$ &$<0$ &$<0$\\

 III(5)  & 0.05  & 1.5  & 1    &$<0$ &$<0$ &$<0$\\

 III(6)  & 0.0001& 100  & 100  &$<0$ &$<0$ &$<0$\\

\end{tabular}
\end{ruledtabular}
\end{table}

In this series of simulations, we keep the surface tension between two droplets constant $\sigma_{gb}=0.05$. The density and viscosity of different components are givne as: $\rho_1/\rho_2=\rho_1/\rho_3=1, \mu_1/\mu_2=\mu_1/\mu_3=1$. The relaxation times are $\tau_\rho=0.1$ and $\tau_\phi=0.3$. Morphology diagram Fig.~\ref{exfig:7} and Table~\ref{Tab:5} show our simulation results of a wide range of surface tension ratios. In Fig.~\ref{exfig:7}, region (I) is composed by double emulsion final morphology. (I-A) and (I-B) separately shows the double emulsion with different outer component due to the spreading factors' difference. Region (II) presents the separate morphology, for which even with a contacting profile initially, two droplets will move to the contrary directions. Region (III) is composed by partially engulfed morphology. The liquid lens or Janus aggregate will appear for equilibrium system. These results are consistent with previous simulation work~\cite{wang2020modelling}.

\subsection{Single rising bubble example}
Single rising bubble process involves the rising bubble dynamics and bubble deformation which are basic problems of many industrial applications such as bubble column reactor and bitumen extraction. Several researchers have investigated the two-phase rising bubble problem before \cite{amaya2010single,amaya2011numerical,hysing2009quantitative}. For this problem, we aim to recover the benchmark morphology which was studied in \cite{hysing2009quantitative}. We initialize a bubble at the bottom of the domain and consider the density ratio $\rho_h/\rho_l=10$, the viscosity ratio $\eta_h/\eta_l=10$, $Bo=10$ and $Ar=35$. We place the bubble with diameters $D/\Delta x=40,80,160$, and $Cn$ can be calculated respectively as $Cn=[ 0.05, 0.025,0.0125]$. The rectangular domain is given as $L \times2L$, where $L/D=2$. Initially, the center of the droplet is placed at $(D,D)$. 
\begin{figure*}[htb!]
  \centering

  \includegraphics[width=\textwidth]{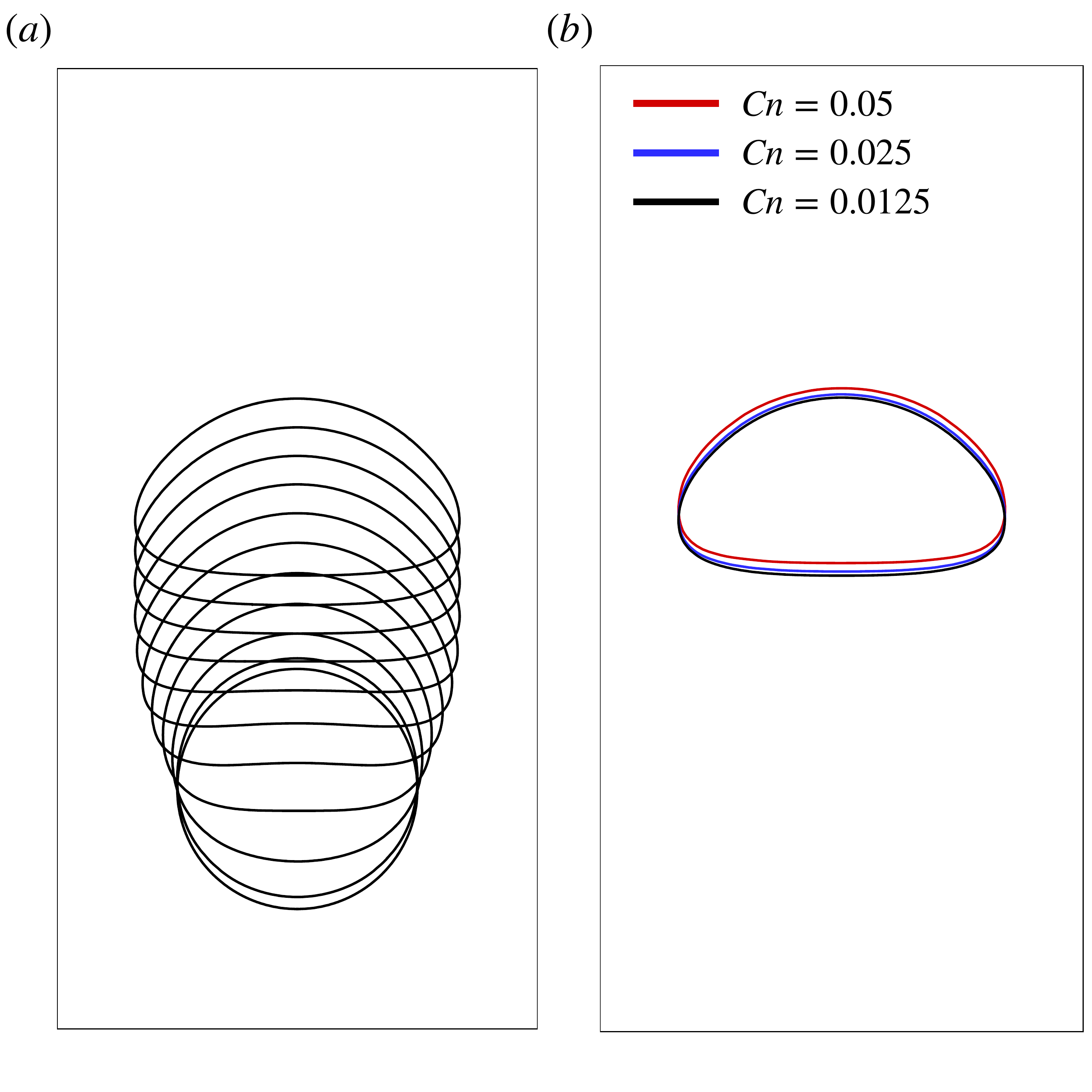}

    \caption{\label{exfig:8}(a) Evolution of rising bubble shapes for $T/t_0=[0-5]$, where time is scaled by $t_0=\sqrt{D/g}$, with $Bo=10, Ar=35$. (b) Convergence test of the bubble shape with $Cn=[0.05,0.025,0.0125]$ at $T/t_0=5$ }

\end{figure*}

\begin{figure*}[htb!]

\centering
  \includegraphics[width=\textwidth]{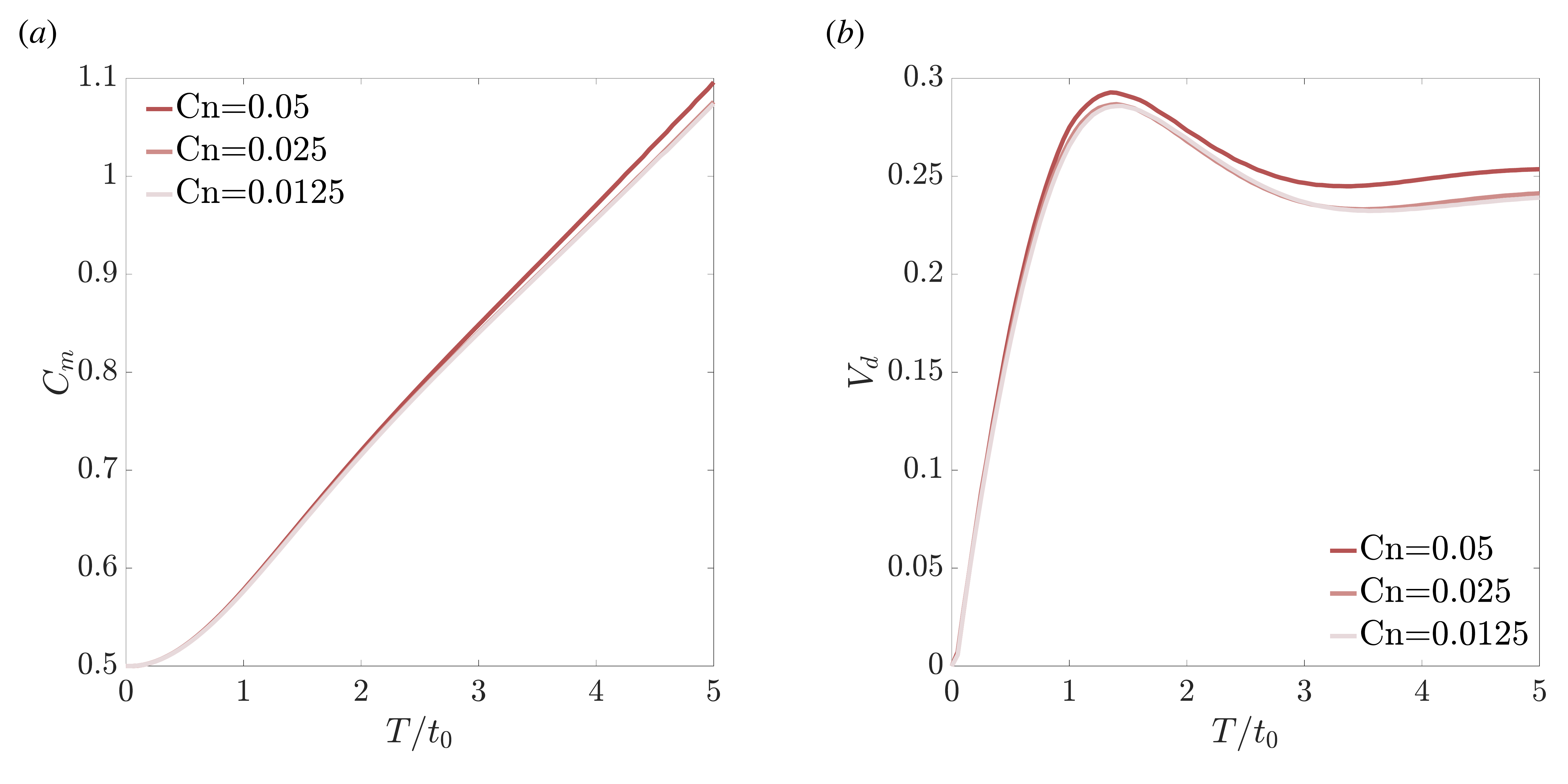}

    \caption{\label{exfig:9} Convergence test of temporal development of (a)mass center of the bubble $C_m$ and (b) average rising velocity $V_d$ which is scaled by $v_0=\sqrt{gD}$ with $Cn=[0.0125,0.025,0.05]$. When $Cn< 0.0375$, the rising velocity and the mass center nearly converge to the same value.}
\end{figure*}

Fig.~\ref{exfig:8}(a) shows the evolution of the rising bubble shapes for $T/t_0=[0-5]$ where $Cn=0.0125$. Fig.~\ref{exfig:8}(b) presents the convergence test of the rising bubbles' morphology with different $Cn$. The bubble shapes of different cases converge as we decrease $Cn$ or increase the number of the grid points. We can then find the temporal development of the mass center of the droplets, $C_m=4\phi_d y/\pi D^2 $  where $\phi_d$ denotes the order parameter for the droplet, and $y$ is the vertical displacement of the droplet according to the axis,  and also the scaled average velocity $V_d$ from Fig.~\ref{exfig:9} (a) and (b). Both the center position and the rising velocity converge to high resolution simulation.

\section{Interaction of a Rising Bubble and a Stationary Droplet}

\begin{table}
\caption{\label{Tab:6}Parameters diagram of ternary flow rising bubble simulation.}
\begin{ruledtabular}
\begin{tabular}{cccc}
 $Bo$ & $Ar$& $Oh$ (Double emulsion) & $Oh$ (Partially engulfed)\\ \hline
    1  & 8  & 0.06    &0.13 \\

    2  & 8  & 0.09    &0.18 \\

    3  & 8  & 0.11    &0.22\\

    4  & 8 & 0.13     &0.25 \\

    5  & 8 & 0.14     &0.28\\

    6  & 8  & 0.15    &0.31 \\

    7  & 8  & 0.17    &0.33\\

    8  & 8  & 0.18    &0.35 \\

\end{tabular}
\end{ruledtabular}
\end{table}

\begin{figure*}[htb!]

  \centering
  \includegraphics[width=\textwidth]{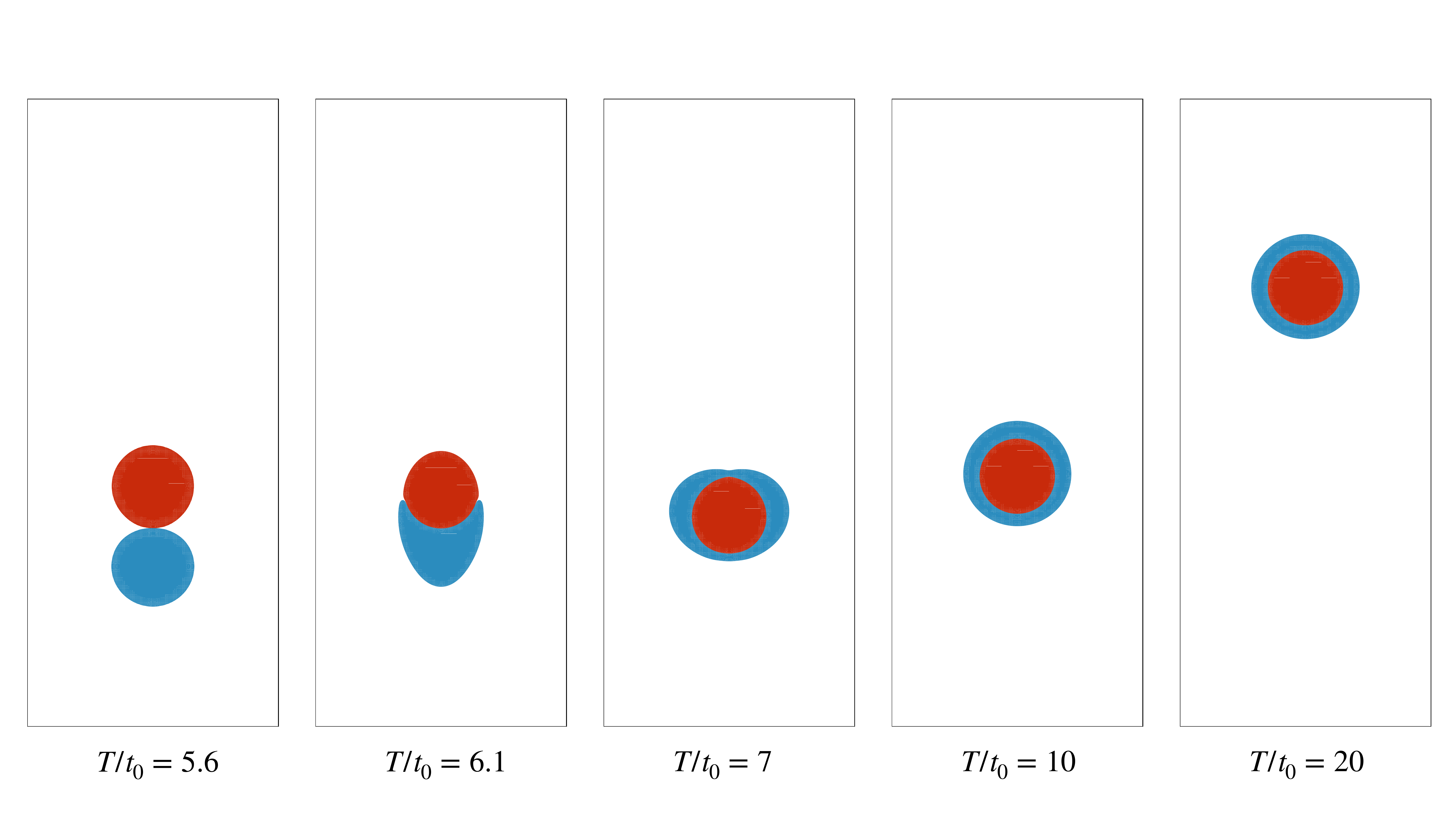}

    \caption{\label{exfig:10}Evolution of the aggregate for $T/t_0=[5.6-20]$, when the surface tension or spreading factors satisfy the condition of the double emulsion morphology under $Bo=1$. The red oil droplet will be fully engulfed by the rising blue bubble at $T/t_0\approx7$.}

\end{figure*}

\begin{figure*}[htb!]
  \centering

  \includegraphics[width=\textwidth]{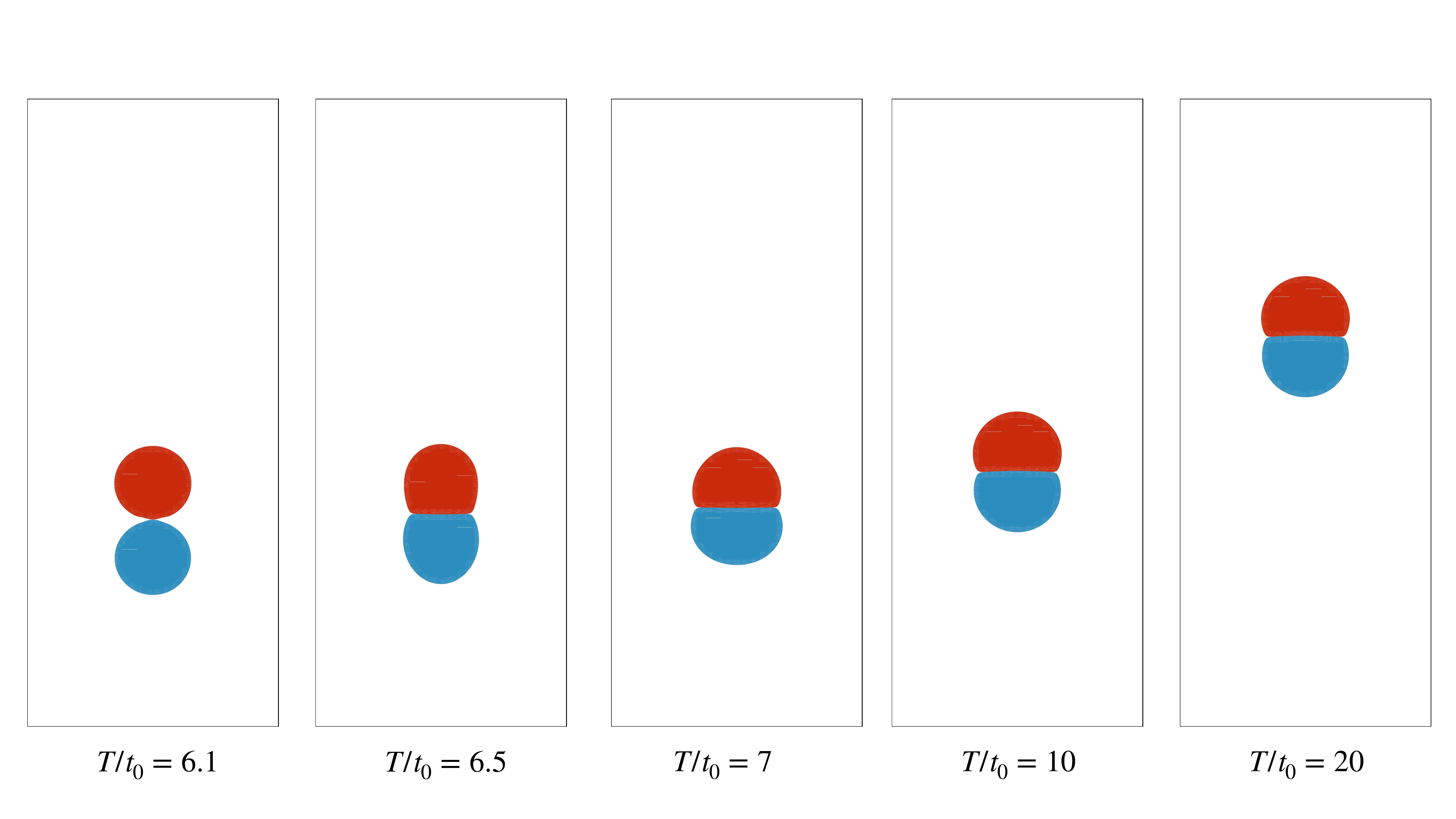}

    \caption{\label{exfig:11}Evolution of the aggregate for $T/t_0=[6.1-20]$, when the surface tension or spreading factors satisfy the condition of the partially engulfed morphology under $Bo=1$. After the blue bubble contacting the red oil droplet, they will maintain this partially engulfed morphology steady, and gradually moving to the top.}

\end{figure*}
\begin{figure*}[htb!]
  \centering

  \includegraphics[width=\textwidth]{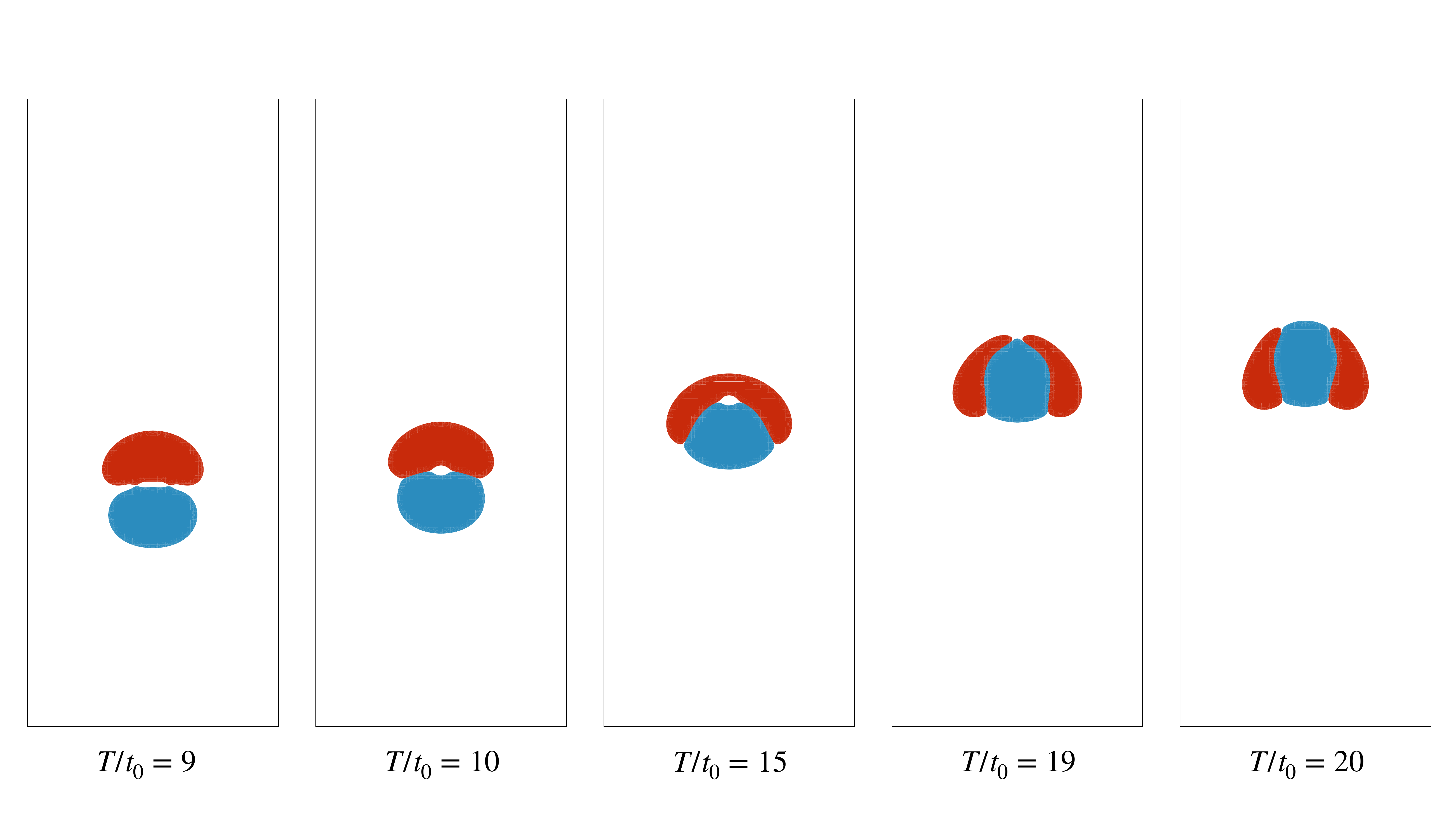}

    \caption{\label{exfig:12}Evolution of the aggregate for $T/t_0=[9-20]$, when the surface tension or spreading factors satisfy the condition of the partially engulfed morphology under $Bo=8$. After the blue bubble contacting the red oil droplet, they first form this partially engulfed morphology. Due to the large rising speed of the aggregate, the oil droplet will break but stick on the rising bubble.}

\end{figure*}

\begin{figure*}[htb!]
\centering
  \includegraphics[width=\textwidth]{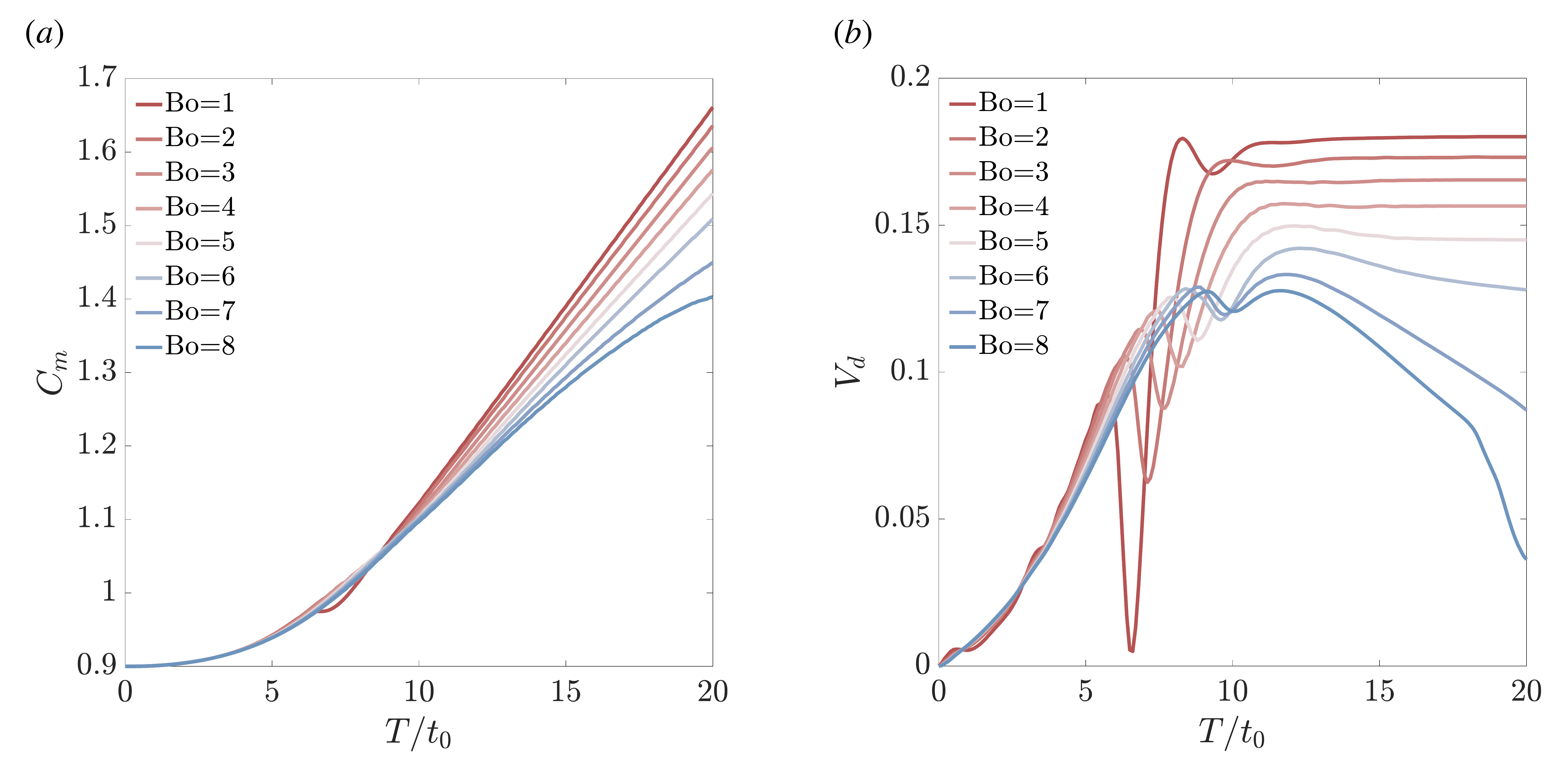}

    \caption{\label{exfig:13} Dynamics of the rising bubble and droplet interaction with partially engulfed morphology. (a) Mass center of the rising bubble $C_m$ development with $Bo=[1-8]$. (b) Scaled average rising velocity $V_d$ of the rising bubble.}
\end{figure*}

\begin{figure*}[htb!]

 \centering
  \includegraphics[width=\textwidth]{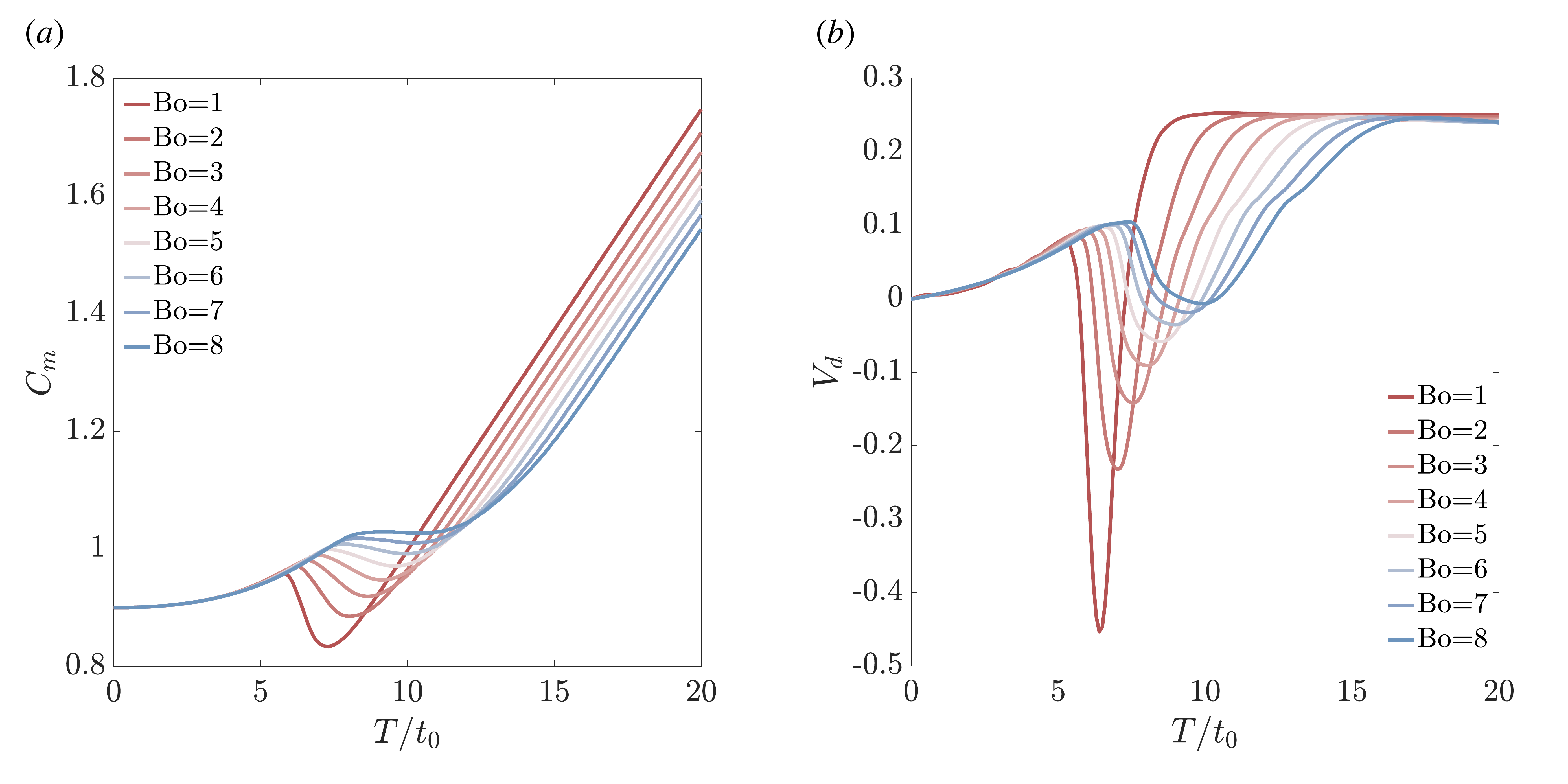}

    \caption{\label{exfig:14} Dynamics of the rising bubble and droplet interaction with double emulsion morphology. (a) Mass center of the rising bubble $C_m$ development with $Bo=[1-8]$. (b) Scaled average rising velocity $V_d$ of the rising bubble.}
\end{figure*}
Based on the single bubble rising test, we set up a ternary flow system to simulate the dynamics of the rising bubble and droplet interaction. The major application of this problem in industry can be found in froth flotation \cite{moosai2003gas,grattoni2003photographic}. The froth flotation extracts the minerals from slurry by assistance with water foam or air bubbles. We model the coalescence and deformation of an air bubble and an oil droplet by imposing a gravitational force. To understand this process, the rising dynamics and morphology change of different size bubbles and oil droplets are needed to be considered. The previous works based on froth flotation concentrate more on the bubble-droplet aggregate forming and ideal cases which include only slow speed and small size bubbles. We here investigate a larger-size bubble and droplet interaction for which we increase $Bo=[1-10]$, and test different spreading factors which provide different perspective to this process.

In a ternary flow, the rising bubble with a lower density will progressively climbs to the top and collides with the oil droplet, which has a density similar to the background fluid. Refer to Fig.~\ref{exfig:7}, when two droplets contact with each other under distinct spreading factors, we can expect different final morphology. Besides, the intensity of surface tension and the viscosity of different components will also affect the interaction dynamics. Hence, $Bo$ and $Ar$ are used to control the rising process, and we utilize $Oh=\frac{\eta_o}{\sqrt{\rho_d\sigma_{dl} D_0}}$ to evaluate the interaction intensity. Here $\rho_d$ is the droplet density, $\sigma_{dl}$ denotes the surface tension between droplet and background liquid, $D_0$ is the diameter of the droplet and $\eta_o$ represents the viscosity of the oil droplet. During the simulation, we keep track of the entire dynamic process by the average rising velocity $V_d=v/v_0$ and the center of the mass of the oil droplet $C_m$ as we did for single droplet rising simulation. We would like to divide the whole rising process into three stages like we introduced earlier: (1) The solitary bubble rises first in a gravitational field; (2) The bubble then makes contact with the top droplet, initiating the interaction; (3) The aggregate rises to the top with a terminal velocity in the final stage.

For the simulation, the diameter of two equal-sized droplets are $D/\Delta x=48$ and the centers are placed at $(20/3,1)\times D$, $(20/3,3)\times D$. The density ratio and viscosity ratio are $\rho_o/\rho_a=100, \rho_o/\rho_l=1$, $\eta_o/\eta_a=100$ and $\eta_o/\eta_l=1$. The interface thickness is $\delta/\Delta x=4$. The no-slip boundary condition is applied to the top and bottom boundaries, while the symmetric boundary condition is applied to the left and right boundaries.

Table~\ref{Tab:6} lists $Bo$, $Ar$, $Oh$ for testing cases. For different $Bo$ with $Ar=8$, a comparison of partially engulfed morphology and double emulsion morphology is proposed. The double emulsion case, $S_2>0$, is excluded since the stable morphology occurs with $Bo\ll1$ which makes the entire rising process inefficient. The time evolution of the rising bubble for different morphological situations with $Bo=1$ is shown in Fig.~\ref{exfig:10} and Fig.~\ref{exfig:11}. Fig.~\ref{exfig:13} and Fig.~\ref{exfig:14} present the rising velocity and center position for different morphology.

We first discuss the case for which the rising bubble will try to partially engulf the oil droplet under this surface tension ratio. As shown in Fig.~\ref{exfig:13}, the cases with small $Bo$ and $Oh$ approach the contact point faster than the cases with large $Bo$ and $Oh$. For inertia regime, $Oh\ll1$, the surface tension force induces a quick contacting or engulfing process for which the viscous resistance hardly affects the dynamics. Under this situation, there exist a severe interaction and a small period engulfing once the rising bubble touches the oil droplet. However, following the interaction, the droplet under a smaller $Bo$ obtains a bigger terminal velocity for the aggregate and continues to rise. It is the morphology comes to affect the system. Situations with large $Bo$ show a velocity decline due to the droplet's distortion. Since compared to gravity, the surface force is still too weak to maintain the shape of this aggregate or partially engulfed morphology, the top droplet splits into two parts for $Bo=8$, as seen in Fig.~\ref{exfig:12}. The center of the droplet exhibits a clear trend that the case with small $Bo$ rises faster during the process.

For double emulsion cases, the initial acceleration and the terminal velocity are quite similar under different $Bo$. From Table~\ref{Tab:6}, we notice that $Oh$ for double emulsion cases are smaller compared to partially engulfed cases. That accounts for the reason why the rising aggregate with double emulsion morphology can maintain this shape unchanged. The surface tension is able to keep the morphology from deforming further, resulting in a comparable drag force. While the cases with small $Oh$ decelerate during the interaction stage, the aggregate of bubble and droplet will keep a even smaller deformation during the rising process compared to large $Bo$ cases.

From our simulation results under $Bo=[1-10]$, the bubble with small $Bo$ which implies a relative smaller diameter in reality maintains its shape and rises fast in both double emulsion and partially engulfed morphology. The rising velocity dominated by gravity is affected by surface force until the completion of the interaction process. However, the intensity of the interaction does not affect rising process too much, the aggregate stability does. Although we expect a quicker rising velocity when $Bo$ is small, the stable double emulsion morphology keeps different size bubbles at same terminal velocity. Among the partially engulfed morphology cases, the drag force breaks up the aggregate and highly influences the rising speed.

\section{Concluding Remarks}
In this article, we presented the simulation work on a rising bubble and a droplet interaction. The conservative phase-field equation was applied as the interface capturing method, and the order parameters were calculated by Lattice Boltzmann equations. In addition, the hydrodynamic properties were calculated by the velocity-pressure-based Lattice Boltzmann equation which recovers the pressure evolution equation and the momentum equation. As for surface force formulation, rather than using potential form formulation, we utilize the CSF formulation to decrease the parasitic currents of the static bubble. Based on convergence tests for both solving single momentum equation and coupled momentum, phase field equation, we argue the CSF can simulate a relative small parasitic currents intensity when curvature is fixed in the simulation. The liquid lens simulation assessed the conservative character and tested the accuracy of the recent method to solve ternary flow system. We verify the current surface tension force applied in this system by the droplet morphology simulation which provide the reference morphology under different spreading factor. 

Based on the single rising bubble simulation, the bubble droplet interaction in a ternary flow is presented. Through the simulations, we learnt that the final rising velocity of the bubble-droplet aggregate highly depends on the morphology stability when $Bo=[1-8]$. We compared the interaction time and the final velocity for various morphology cases. A smaller $Bo$ and $Oh$ resulted in a faster interaction process with a higher interaction intensity. We also detected a higher terminal velocity with a smaller $Bo$ for partially engulfed morphology. Due to slight distortion, the cases of double emulsion morphology achieved a similar terminal velocity among different $Bo$.
\section*{Acknowledgement}
This research was supported by the National Science Founda- tion under Grant No. 1743794, PIRE: Investigation of Multi-Scale, Multi-Phase Phenomena in Complex Fluids for the Energy Indus- tries.
\clearpage
\appendix
\section{Chapman Enskog Analysis}
We present a Chapman-Enskog Analysis based on the Discrete Boltzmann equation in this section. The Discrete Boltzmann equation is given as Eq.\ref{Ceq:11}:
\begin{equation}\label{Ap:1}
  \frac{\partial g_\alpha }{\partial t}+\boldsymbol{e}_\alpha \cdot \nabla g_\alpha=-\frac{g_\alpha-g^{eq}}{\lambda}+F_\alpha
\end{equation}
We consider $\delta t$ to be the small parameter in this case, thus the fundamental expansions based on $\delta t$ for distribution function and the time derivative are expressed as follows:
\begin{equation}\label{Ap:2}
  g_\alpha(\boldsymbol{x},t) =g_\alpha^{eq}(\boldsymbol{x},t)+\delta t g_\alpha^{(1)}(\boldsymbol{x},t)+\delta t^2g_\alpha^{(2)}(\boldsymbol{x},t)
\end{equation}
\begin{equation}\label{Ap:3}
\partial_t=\partial_{t0} +\delta t \partial_{t1}
\end{equation}

We obtain the $\delta t$ order equation after some calculations:
\begin{equation}\label{Ap:4}
  \frac{\partial g_\alpha^{eq}}{\partial t_0}=\frac{g_\alpha^{(1)}}{\tau}+F_\alpha
\end{equation}
and the $\delta t^2$ order equation can be expressed as:
\begin{equation}\label{Ap:5}
  \frac{\partial g_\alpha^{eq}}{\partial t_1}+(\partial t_0+\boldsymbol{e}_\alpha\cdot\nabla)g_\alpha^{(1)}=\frac{g_\alpha^{(2)}}{\tau}
\end{equation}
Because we aim to recover the partial differential equation, we complete summation of $O(\delta t)+\delta t O(\delta t^2)$ and cut off the high order terms. The equation can then be derived as:
\begin{equation}\label{Ap:6}
  \frac{\partial g_\alpha^{eq}}{\partial t}+\boldsymbol{e}_\alpha\cdot\nabla g_\alpha^{eq}+\delta t(\partial_{t0}+\boldsymbol{e}_\alpha\cdot\nabla)g_\alpha^{(1)}=-\frac{1}{\lambda}(g_\alpha-g_\alpha^{eq})+F_\alpha
\end{equation}
We then restrict the moments of the equilibrium distribution to obtain the macroscopic value from the distribution function:
\begin{equation}\label{Ap:7}
  \sum_{\alpha}g_\alpha^{eq}=\bar{p}
\end{equation}
\begin{equation}\label{Ap:8}
  \sum_{\alpha}g_\alpha^{eq}\boldsymbol{e}_\alpha=\boldsymbol{u}c_s^2
\end{equation}
\begin{equation}\label{Ap:9}
 \sum_{\alpha}g_\alpha^{eq}\boldsymbol{e}_\alpha \boldsymbol{e}_\alpha=\boldsymbol{u}\boldsymbol{u}c_s^2+\bar{p}c_s^2
\end{equation}
and the moments of the source term:
\begin{equation}\label{Ap:10}
  \sum_\alpha F_\alpha=-\boldsymbol{u}\cdot\nabla\bar{p}
\end{equation}
\begin{equation}\label{Ap:11}
\sum_\alpha F_\alpha \boldsymbol{e}_\alpha =\frac{c_s^2}{\rho}\left(-\nabla P+\rho\nabla\bar{p}+\nu(\nabla\boldsymbol{u}+\nabla\boldsymbol{u}^T)\nabla\rho+\boldsymbol{F}_s+\boldsymbol{F}_b\right)
  \end{equation}
\begin{equation}\label{Ap:12}
\sum_\alpha F_\alpha \boldsymbol{e}_\alpha\boldsymbol{e}_\alpha =c_s^2\boldsymbol{u}\cdot\nabla\bar{p}
  \end{equation}

Under these conditions, the following equations can be deducted from the zeroth and the first moments of Eq.\ref{Ap:6} :
\begin{equation}\label{Ap:13}
  \frac{\partial \bar{p}}{\partial t}+\nabla\cdot\boldsymbol{u}c_s^2+\boldsymbol{u}\cdot\nabla \bar{p}=0
\end{equation}
\begin{equation}\label{Ap:14}
  \frac{\partial\boldsymbol{u}}{\partial t}+\nabla\cdot\boldsymbol{u}\boldsymbol{u}+\frac{\delta t}{c_s^2}\nabla\cdot\Pi^{(1)}=-\frac{1}{\rho}\nabla P+\frac{\nu}{\rho}(\nabla\boldsymbol{u}+\nabla\boldsymbol{u}^T)\nabla\rho+\frac{F_s}{\rho}+\frac{F_b}{\rho}
\end{equation}

We can further decompose Eq.\ref{Ap:13} into continuity equation and pressure evolution equation. When we consider the kinematic viscosity $\nu=\tau c_s^2\delta t$, Eq.\ref{Ap:14} becomes:
\begin{equation}\label{Ap:15}
  \frac{\partial\boldsymbol{u}}{\partial t}+\nabla\cdot\boldsymbol{u}\boldsymbol{u}=-\frac{1}{\rho}\nabla P+\frac{1}{\rho}\nabla\cdot\eta(\nabla\boldsymbol{u}+\nabla\boldsymbol{u}^T)+\frac{F_s}{\rho}+\frac{F_b}{\rho}
\end{equation}
Where $\eta=\nu\rho$ is the dynamic viscosity

In the final, the governing equations, Eq.\ref{Ap:14} and Eq.\ref{Ap:15}, are retrieved from the Discrete Boltzmann equation \ref{Ap:1}.

\section{Contact angle calculation}
\begin{figure}[htb!]

  \centering
  \includegraphics[width=0.5\textwidth]{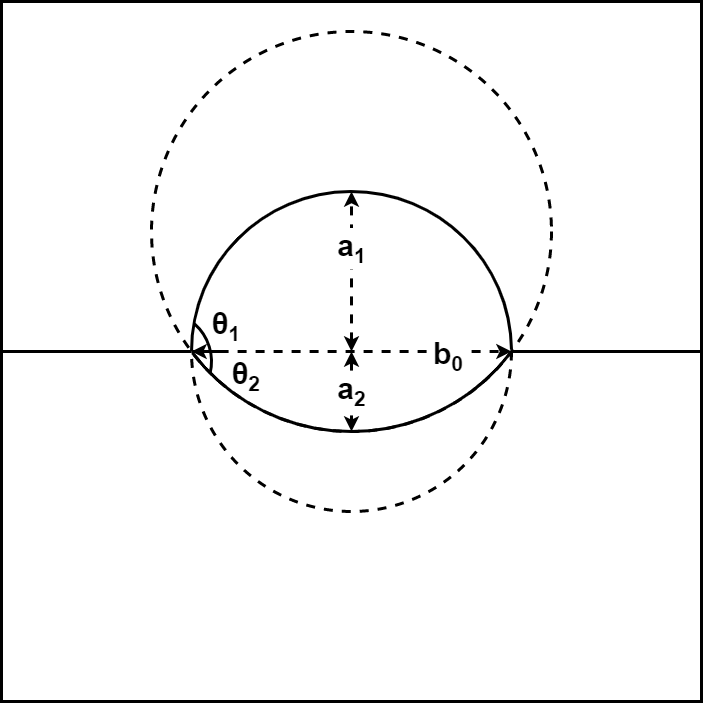}

    \caption{\label{exfig:3}The schematic of spreading of a liquid lens.}

\end{figure}
In this section, we show the method how to obtain the contact angle from the liquid lens simulation. An stationary liquid lens schematic is shown in FIG.\ref{exfig:3}, where the droplet achieves equilibrium state between two background fluids. In this figure, $b_0$ is the length between two three-phase contact points. $a_1$ and $a_2$ are the distances between the droplet top and bottom and the two-phase contact line. $b_0$, $a_1$, $a_2$ can be obtained by the order parameter $\phi=0.5$. the upper contact angle and lower contact angle $\theta_1$ and $\theta_2$ can then be calculated as: $\theta^{eq}_i=180^\circ-2arctan(\frac{b_0}{2a_i})$. Finally, we can calculate the contact angle as $\theta=\theta_1+\theta_2$.

\bibliography{apssamp}

\providecommand{\noopsort}[1]{}\providecommand{\singleletter}[1]{#1}%
\begin{thebibliography}{45}%
\makeatletter
\providecommand \@ifxundefined [1]{%
 \@ifx{#1\undefined}
}%
\providecommand \@ifnum [1]{%
 \ifnum #1\expandafter \@firstoftwo
 \else \expandafter \@secondoftwo
 \fi
}%
\providecommand \@ifx [1]{%
 \ifx #1\expandafter \@firstoftwo
 \else \expandafter \@secondoftwo
 \fi
}%
\providecommand \natexlab [1]{#1}%
\providecommand \enquote  [1]{``#1''}%
\providecommand \bibnamefont  [1]{#1}%
\providecommand \bibfnamefont [1]{#1}%
\providecommand \citenamefont [1]{#1}%
\providecommand \href@noop [0]{\@secondoftwo}%
\providecommand \href [0]{\begingroup \@sanitize@url \@href}%
\providecommand \@href[1]{\@@startlink{#1}\@@href}%
\providecommand \@@href[1]{\endgroup#1\@@endlink}%
\providecommand \@sanitize@url [0]{\catcode `\\12\catcode `\$12\catcode
  `\&12\catcode `\#12\catcode `\^12\catcode `\_12\catcode `\%12\relax}%
\providecommand \@@startlink[1]{}%
\providecommand \@@endlink[0]{}%
\providecommand \url  [0]{\begingroup\@sanitize@url \@url }%
\providecommand \@url [1]{\endgroup\@href {#1}{\urlprefix }}%
\providecommand \urlprefix  [0]{URL }%
\providecommand \Eprint [0]{\href }%
\providecommand \doibase [0]{https://doi.org/}%
\providecommand \selectlanguage [0]{\@gobble}%
\providecommand \bibinfo  [0]{\@secondoftwo}%
\providecommand \bibfield  [0]{\@secondoftwo}%
\providecommand \translation [1]{[#1]}%
\providecommand \BibitemOpen [0]{}%
\providecommand \bibitemStop [0]{}%
\providecommand \bibitemNoStop [0]{.\EOS\space}%
\providecommand \EOS [0]{\spacefactor3000\relax}%
\providecommand \BibitemShut  [1]{\csname bibitem#1\endcsname}%
\let\auto@bib@innerbib\@empty
\bibitem [{\citenamefont {Moosai}\ and\ \citenamefont
  {Dawe}(2003)}]{moosai2003gas}%
  \BibitemOpen
  \bibfield  {author} {\bibinfo {author} {\bibfnamefont {R.}~\bibnamefont
  {Moosai}}\ and\ \bibinfo {author} {\bibfnamefont {R.~A.}\ \bibnamefont
  {Dawe}},\ }\bibfield  {title} {\bibinfo {title} {Gas attachment of oil
  droplets for gas flotation for oily wastewater cleanup},\ }\href@noop {}
  {\bibfield  {journal} {\bibinfo  {journal} {Separation and purification
  technology}\ }\textbf {\bibinfo {volume} {33}},\ \bibinfo {pages} {303}
  (\bibinfo {year} {2003})}\BibitemShut {NoStop}%
\bibitem [{\citenamefont {Saththasivam}\ \emph {et~al.}(2016)\citenamefont
  {Saththasivam}, \citenamefont {Loganathan},\ and\ \citenamefont
  {Sarp}}]{saththasivam2016overview}%
  \BibitemOpen
  \bibfield  {author} {\bibinfo {author} {\bibfnamefont {J.}~\bibnamefont
  {Saththasivam}}, \bibinfo {author} {\bibfnamefont {K.}~\bibnamefont
  {Loganathan}},\ and\ \bibinfo {author} {\bibfnamefont {S.}~\bibnamefont
  {Sarp}},\ }\bibfield  {title} {\bibinfo {title} {An overview of oil--water
  separation using gas flotation systems},\ }\href@noop {} {\bibfield
  {journal} {\bibinfo  {journal} {Chemosphere}\ }\textbf {\bibinfo {volume}
  {144}},\ \bibinfo {pages} {671} (\bibinfo {year} {2016})}\BibitemShut
  {NoStop}%
\bibitem [{\citenamefont {Grattoni}\ \emph {et~al.}(2003)\citenamefont
  {Grattoni}, \citenamefont {Moosai},\ and\ \citenamefont
  {Dawe}}]{grattoni2003photographic}%
  \BibitemOpen
  \bibfield  {author} {\bibinfo {author} {\bibfnamefont {C.}~\bibnamefont
  {Grattoni}}, \bibinfo {author} {\bibfnamefont {R.}~\bibnamefont {Moosai}},\
  and\ \bibinfo {author} {\bibfnamefont {R.~A.}\ \bibnamefont {Dawe}},\
  }\bibfield  {title} {\bibinfo {title} {Photographic observations showing
  spreading and non-spreading of oil on gas bubbles of relevance to gas
  flotation for oily wastewater cleanup},\ }\href@noop {} {\bibfield  {journal}
  {\bibinfo  {journal} {Colloids and Surfaces A: Physicochemical and
  Engineering Aspects}\ }\textbf {\bibinfo {volume} {214}},\ \bibinfo {pages}
  {151} (\bibinfo {year} {2003})}\BibitemShut {NoStop}%
\bibitem [{\citenamefont {Amaya-Bower}\ and\ \citenamefont
  {Lee}(2010)}]{amaya2010single}%
  \BibitemOpen
  \bibfield  {author} {\bibinfo {author} {\bibfnamefont {L.}~\bibnamefont
  {Amaya-Bower}}\ and\ \bibinfo {author} {\bibfnamefont {T.}~\bibnamefont
  {Lee}},\ }\bibfield  {title} {\bibinfo {title} {Single bubble rising dynamics
  for moderate reynolds number using lattice boltzmann method},\ }\href@noop {}
  {\bibfield  {journal} {\bibinfo  {journal} {Computers \& Fluids}\ }\textbf
  {\bibinfo {volume} {39}},\ \bibinfo {pages} {1191} (\bibinfo {year}
  {2010})}\BibitemShut {NoStop}%
\bibitem [{\citenamefont {Amaya-Bower}\ and\ \citenamefont
  {Lee}(2011)}]{amaya2011numerical}%
  \BibitemOpen
  \bibfield  {author} {\bibinfo {author} {\bibfnamefont {L.}~\bibnamefont
  {Amaya-Bower}}\ and\ \bibinfo {author} {\bibfnamefont {T.}~\bibnamefont
  {Lee}},\ }\bibfield  {title} {\bibinfo {title} {Numerical simulation of
  single bubble rising in vertical and inclined square channel using lattice
  boltzmann method},\ }\href@noop {} {\bibfield  {journal} {\bibinfo  {journal}
  {Chemical Engineering Science}\ }\textbf {\bibinfo {volume} {66}},\ \bibinfo
  {pages} {935} (\bibinfo {year} {2011})}\BibitemShut {NoStop}%
\bibitem [{\citenamefont {Pannacci}\ \emph {et~al.}(2008)\citenamefont
  {Pannacci}, \citenamefont {Bruus}, \citenamefont {Bartolo}, \citenamefont
  {Etchart}, \citenamefont {Lockhart}, \citenamefont {Hennequin}, \citenamefont
  {Willaime},\ and\ \citenamefont {Tabeling}}]{pannacci2008equilibrium}%
  \BibitemOpen
  \bibfield  {author} {\bibinfo {author} {\bibfnamefont {N.}~\bibnamefont
  {Pannacci}}, \bibinfo {author} {\bibfnamefont {H.}~\bibnamefont {Bruus}},
  \bibinfo {author} {\bibfnamefont {D.}~\bibnamefont {Bartolo}}, \bibinfo
  {author} {\bibfnamefont {I.}~\bibnamefont {Etchart}}, \bibinfo {author}
  {\bibfnamefont {T.}~\bibnamefont {Lockhart}}, \bibinfo {author}
  {\bibfnamefont {Y.}~\bibnamefont {Hennequin}}, \bibinfo {author}
  {\bibfnamefont {H.}~\bibnamefont {Willaime}},\ and\ \bibinfo {author}
  {\bibfnamefont {P.}~\bibnamefont {Tabeling}},\ }\bibfield  {title} {\bibinfo
  {title} {Equilibrium and nonequilibrium states in microfluidic double
  emulsions},\ }\href@noop {} {\bibfield  {journal} {\bibinfo  {journal}
  {Physical review letters}\ }\textbf {\bibinfo {volume} {101}},\ \bibinfo
  {pages} {164502} (\bibinfo {year} {2008})}\BibitemShut {NoStop}%
\bibitem [{\citenamefont {Guzowski}\ \emph {et~al.}(2012)\citenamefont
  {Guzowski}, \citenamefont {Korczyk}, \citenamefont {Jakiela},\ and\
  \citenamefont {Garstecki}}]{guzowski2012structure}%
  \BibitemOpen
  \bibfield  {author} {\bibinfo {author} {\bibfnamefont {J.}~\bibnamefont
  {Guzowski}}, \bibinfo {author} {\bibfnamefont {P.~M.}\ \bibnamefont
  {Korczyk}}, \bibinfo {author} {\bibfnamefont {S.}~\bibnamefont {Jakiela}},\
  and\ \bibinfo {author} {\bibfnamefont {P.}~\bibnamefont {Garstecki}},\
  }\bibfield  {title} {\bibinfo {title} {The structure and stability of
  multiple micro-droplets},\ }\href@noop {} {\bibfield  {journal} {\bibinfo
  {journal} {Soft Matter}\ }\textbf {\bibinfo {volume} {8}},\ \bibinfo {pages}
  {7269} (\bibinfo {year} {2012})}\BibitemShut {NoStop}%
\bibitem [{\citenamefont {Jacqmin}(1996)}]{jacqmin1996energy}%
  \BibitemOpen
  \bibfield  {author} {\bibinfo {author} {\bibfnamefont {D.}~\bibnamefont
  {Jacqmin}},\ }\bibfield  {title} {\bibinfo {title} {An energy approach to the
  continuum surface tension method},\ }in\ \href@noop {} {\emph {\bibinfo
  {booktitle} {34th Aerospace sciences meeting and exhibit}}}\ (\bibinfo {year}
  {1996})\ p.\ \bibinfo {pages} {858}\BibitemShut {NoStop}%
\bibitem [{\citenamefont {Baroudi}\ and\ \citenamefont
  {Lee}(2021)}]{baroudi2021simulation}%
  \BibitemOpen
  \bibfield  {author} {\bibinfo {author} {\bibfnamefont {L.}~\bibnamefont
  {Baroudi}}\ and\ \bibinfo {author} {\bibfnamefont {T.}~\bibnamefont {Lee}},\
  }\bibfield  {title} {\bibinfo {title} {Simulation of a bubble rising at high
  reynolds number with mass-conserving finite element lattice boltzmann
  method},\ }\href@noop {} {\bibfield  {journal} {\bibinfo  {journal}
  {Computers \& Fluids}\ }\textbf {\bibinfo {volume} {220}},\ \bibinfo {pages}
  {104883} (\bibinfo {year} {2021})}\BibitemShut {NoStop}%
\bibitem [{\citenamefont {Cahn}(1959)}]{cahn1959free}%
  \BibitemOpen
  \bibfield  {author} {\bibinfo {author} {\bibfnamefont {J.~W.}\ \bibnamefont
  {Cahn}},\ }\bibfield  {title} {\bibinfo {title} {Free energy of a nonuniform
  system. ii. thermodynamic basis},\ }\href@noop {} {\bibfield  {journal}
  {\bibinfo  {journal} {The Journal of chemical physics}\ }\textbf {\bibinfo
  {volume} {30}},\ \bibinfo {pages} {1121} (\bibinfo {year}
  {1959})}\BibitemShut {NoStop}%
\bibitem [{\citenamefont {Allen}\ and\ \citenamefont
  {Cahn}(1979)}]{allen1979microscopic}%
  \BibitemOpen
  \bibfield  {author} {\bibinfo {author} {\bibfnamefont {S.~M.}\ \bibnamefont
  {Allen}}\ and\ \bibinfo {author} {\bibfnamefont {J.~W.}\ \bibnamefont
  {Cahn}},\ }\bibfield  {title} {\bibinfo {title} {A microscopic theory for
  antiphase boundary motion and its application to antiphase domain
  coarsening},\ }\href@noop {} {\bibfield  {journal} {\bibinfo  {journal} {Acta
  metallurgica}\ }\textbf {\bibinfo {volume} {27}},\ \bibinfo {pages} {1085}
  (\bibinfo {year} {1979})}\BibitemShut {NoStop}%
\bibitem [{\citenamefont {Kim}(2005)}]{kim2005continuous}%
  \BibitemOpen
  \bibfield  {author} {\bibinfo {author} {\bibfnamefont {J.}~\bibnamefont
  {Kim}},\ }\bibfield  {title} {\bibinfo {title} {A continuous surface tension
  force formulation for diffuse-interface models},\ }\href@noop {} {\bibfield
  {journal} {\bibinfo  {journal} {Journal of Computational Physics}\ }\textbf
  {\bibinfo {volume} {204}},\ \bibinfo {pages} {784} (\bibinfo {year}
  {2005})}\BibitemShut {NoStop}%
\bibitem [{\citenamefont {Lee}\ and\ \citenamefont
  {Liu}(2010)}]{lee2010lattice}%
  \BibitemOpen
  \bibfield  {author} {\bibinfo {author} {\bibfnamefont {T.}~\bibnamefont
  {Lee}}\ and\ \bibinfo {author} {\bibfnamefont {L.}~\bibnamefont {Liu}},\
  }\bibfield  {title} {\bibinfo {title} {Lattice boltzmann simulations of
  micron-scale drop impact on dry surfaces},\ }\href@noop {} {\bibfield
  {journal} {\bibinfo  {journal} {Journal of Computational Physics}\ }\textbf
  {\bibinfo {volume} {229}},\ \bibinfo {pages} {8045} (\bibinfo {year}
  {2010})}\BibitemShut {NoStop}%
\bibitem [{\citenamefont {Lee}\ and\ \citenamefont
  {Lin}(2005)}]{lee2005stable}%
  \BibitemOpen
  \bibfield  {author} {\bibinfo {author} {\bibfnamefont {T.}~\bibnamefont
  {Lee}}\ and\ \bibinfo {author} {\bibfnamefont {C.-L.}\ \bibnamefont {Lin}},\
  }\bibfield  {title} {\bibinfo {title} {A stable discretization of the lattice
  boltzmann equation for simulation of incompressible two-phase flows at high
  density ratio},\ }\href@noop {} {\bibfield  {journal} {\bibinfo  {journal}
  {Journal of Computational Physics}\ }\textbf {\bibinfo {volume} {206}},\
  \bibinfo {pages} {16} (\bibinfo {year} {2005})}\BibitemShut {NoStop}%
\bibitem [{\citenamefont {Abadi}\ \emph
  {et~al.}(2018{\natexlab{a}})\citenamefont {Abadi}, \citenamefont {Fakhari},\
  and\ \citenamefont {Rahimian}}]{abadi2018numerical}%
  \BibitemOpen
  \bibfield  {author} {\bibinfo {author} {\bibfnamefont {R.~H.~H.}\
  \bibnamefont {Abadi}}, \bibinfo {author} {\bibfnamefont {A.}~\bibnamefont
  {Fakhari}},\ and\ \bibinfo {author} {\bibfnamefont {M.~H.}\ \bibnamefont
  {Rahimian}},\ }\bibfield  {title} {\bibinfo {title} {Numerical simulation of
  three-component multiphase flows at high density and viscosity ratios using
  lattice boltzmann methods},\ }\href@noop {} {\bibfield  {journal} {\bibinfo
  {journal} {Physical Review E}\ }\textbf {\bibinfo {volume} {97}},\ \bibinfo
  {pages} {033312} (\bibinfo {year} {2018}{\natexlab{a}})}\BibitemShut
  {NoStop}%
\bibitem [{\citenamefont {Sun}\ and\ \citenamefont
  {Beckermann}(2007)}]{sun2007sharp}%
  \BibitemOpen
  \bibfield  {author} {\bibinfo {author} {\bibfnamefont {Y.}~\bibnamefont
  {Sun}}\ and\ \bibinfo {author} {\bibfnamefont {C.}~\bibnamefont
  {Beckermann}},\ }\bibfield  {title} {\bibinfo {title} {Sharp interface
  tracking using the phase-field equation},\ }\href@noop {} {\bibfield
  {journal} {\bibinfo  {journal} {Journal of Computational Physics}\ }\textbf
  {\bibinfo {volume} {220}},\ \bibinfo {pages} {626} (\bibinfo {year}
  {2007})}\BibitemShut {NoStop}%
\bibitem [{\citenamefont {Yue}\ \emph {et~al.}(2007)\citenamefont {Yue},
  \citenamefont {Zhou},\ and\ \citenamefont {Feng}}]{yue2007spontaneous}%
  \BibitemOpen
  \bibfield  {author} {\bibinfo {author} {\bibfnamefont {P.}~\bibnamefont
  {Yue}}, \bibinfo {author} {\bibfnamefont {C.}~\bibnamefont {Zhou}},\ and\
  \bibinfo {author} {\bibfnamefont {J.~J.}\ \bibnamefont {Feng}},\ }\bibfield
  {title} {\bibinfo {title} {Spontaneous shrinkage of drops and mass
  conservation in phase-field simulations},\ }\href@noop {} {\bibfield
  {journal} {\bibinfo  {journal} {Journal of Computational Physics}\ }\textbf
  {\bibinfo {volume} {223}},\ \bibinfo {pages} {1} (\bibinfo {year}
  {2007})}\BibitemShut {NoStop}%
\bibitem [{\citenamefont {Zheng}\ \emph {et~al.}(2014)\citenamefont {Zheng},
  \citenamefont {Lee}, \citenamefont {Guo},\ and\ \citenamefont
  {Rumschitzki}}]{zheng2014shrinkage}%
  \BibitemOpen
  \bibfield  {author} {\bibinfo {author} {\bibfnamefont {L.}~\bibnamefont
  {Zheng}}, \bibinfo {author} {\bibfnamefont {T.}~\bibnamefont {Lee}}, \bibinfo
  {author} {\bibfnamefont {Z.}~\bibnamefont {Guo}},\ and\ \bibinfo {author}
  {\bibfnamefont {D.}~\bibnamefont {Rumschitzki}},\ }\bibfield  {title}
  {\bibinfo {title} {Shrinkage of bubbles and drops in the lattice boltzmann
  equation method for nonideal gases},\ }\href@noop {} {\bibfield  {journal}
  {\bibinfo  {journal} {Physical Review E}\ }\textbf {\bibinfo {volume} {89}},\
  \bibinfo {pages} {033302} (\bibinfo {year} {2014})}\BibitemShut {NoStop}%
\bibitem [{\citenamefont {Folch}\ \emph {et~al.}(1999)\citenamefont {Folch},
  \citenamefont {Casademunt}, \citenamefont {Hern{\'a}ndez-Machado},\ and\
  \citenamefont {Ram{\'\i}rez-Piscina}}]{folch1999phase}%
  \BibitemOpen
  \bibfield  {author} {\bibinfo {author} {\bibfnamefont {R.}~\bibnamefont
  {Folch}}, \bibinfo {author} {\bibfnamefont {J.}~\bibnamefont {Casademunt}},
  \bibinfo {author} {\bibfnamefont {A.}~\bibnamefont {Hern{\'a}ndez-Machado}},\
  and\ \bibinfo {author} {\bibfnamefont {L.}~\bibnamefont
  {Ram{\'\i}rez-Piscina}},\ }\bibfield  {title} {\bibinfo {title} {Phase-field
  model for hele-shaw flows with arbitrary viscosity contrast. ii. numerical
  study},\ }\href@noop {} {\bibfield  {journal} {\bibinfo  {journal} {Physical
  Review E}\ }\textbf {\bibinfo {volume} {60}},\ \bibinfo {pages} {1734}
  (\bibinfo {year} {1999})}\BibitemShut {NoStop}%
\bibitem [{\citenamefont {Chiu}\ and\ \citenamefont
  {Lin}(2011)}]{chiu2011conservative}%
  \BibitemOpen
  \bibfield  {author} {\bibinfo {author} {\bibfnamefont {P.-H.}\ \bibnamefont
  {Chiu}}\ and\ \bibinfo {author} {\bibfnamefont {Y.-T.}\ \bibnamefont {Lin}},\
  }\bibfield  {title} {\bibinfo {title} {A conservative phase field method for
  solving incompressible two-phase flows},\ }\href@noop {} {\bibfield
  {journal} {\bibinfo  {journal} {Journal of Computational Physics}\ }\textbf
  {\bibinfo {volume} {230}},\ \bibinfo {pages} {185} (\bibinfo {year}
  {2011})}\BibitemShut {NoStop}%
\bibitem [{\citenamefont {Geier}\ \emph {et~al.}(2015)\citenamefont {Geier},
  \citenamefont {Fakhari},\ and\ \citenamefont {Lee}}]{geier2015conservative}%
  \BibitemOpen
  \bibfield  {author} {\bibinfo {author} {\bibfnamefont {M.}~\bibnamefont
  {Geier}}, \bibinfo {author} {\bibfnamefont {A.}~\bibnamefont {Fakhari}},\
  and\ \bibinfo {author} {\bibfnamefont {T.}~\bibnamefont {Lee}},\ }\bibfield
  {title} {\bibinfo {title} {Conservative phase-field lattice boltzmann model
  for interface tracking equation},\ }\href@noop {} {\bibfield  {journal}
  {\bibinfo  {journal} {Physical Review E}\ }\textbf {\bibinfo {volume} {91}},\
  \bibinfo {pages} {063309} (\bibinfo {year} {2015})}\BibitemShut {NoStop}%
\bibitem [{\citenamefont {Abadi}\ \emph
  {et~al.}(2018{\natexlab{b}})\citenamefont {Abadi}, \citenamefont {Rahimian},\
  and\ \citenamefont {Fakhari}}]{abadi2018conservative}%
  \BibitemOpen
  \bibfield  {author} {\bibinfo {author} {\bibfnamefont {R.~H.~H.}\
  \bibnamefont {Abadi}}, \bibinfo {author} {\bibfnamefont {M.~H.}\ \bibnamefont
  {Rahimian}},\ and\ \bibinfo {author} {\bibfnamefont {A.}~\bibnamefont
  {Fakhari}},\ }\bibfield  {title} {\bibinfo {title} {Conservative phase-field
  lattice-boltzmann model for ternary fluids},\ }\href@noop {} {\bibfield
  {journal} {\bibinfo  {journal} {Journal of Computational Physics}\ }\textbf
  {\bibinfo {volume} {374}},\ \bibinfo {pages} {668} (\bibinfo {year}
  {2018}{\natexlab{b}})}\BibitemShut {NoStop}%
\bibitem [{\citenamefont {Aihara}\ \emph {et~al.}(2019)\citenamefont {Aihara},
  \citenamefont {Takaki},\ and\ \citenamefont {Takada}}]{aihara2019multi}%
  \BibitemOpen
  \bibfield  {author} {\bibinfo {author} {\bibfnamefont {S.}~\bibnamefont
  {Aihara}}, \bibinfo {author} {\bibfnamefont {T.}~\bibnamefont {Takaki}},\
  and\ \bibinfo {author} {\bibfnamefont {N.}~\bibnamefont {Takada}},\
  }\bibfield  {title} {\bibinfo {title} {Multi-phase-field modeling using a
  conservative allen--cahn equation for multiphase flow},\ }\href@noop {}
  {\bibfield  {journal} {\bibinfo  {journal} {Computers \& Fluids}\ }\textbf
  {\bibinfo {volume} {178}},\ \bibinfo {pages} {141} (\bibinfo {year}
  {2019})}\BibitemShut {NoStop}%
\bibitem [{\citenamefont {Zheng}\ \emph {et~al.}(2020)\citenamefont {Zheng},
  \citenamefont {Zheng},\ and\ \citenamefont {Zhai}}]{zheng2020multiphase}%
  \BibitemOpen
  \bibfield  {author} {\bibinfo {author} {\bibfnamefont {L.}~\bibnamefont
  {Zheng}}, \bibinfo {author} {\bibfnamefont {S.}~\bibnamefont {Zheng}},\ and\
  \bibinfo {author} {\bibfnamefont {Q.}~\bibnamefont {Zhai}},\ }\bibfield
  {title} {\bibinfo {title} {Multiphase flows of n immiscible incompressible
  fluids: Conservative allen-cahn equation and lattice boltzmann equation
  method},\ }\href@noop {} {\bibfield  {journal} {\bibinfo  {journal} {Physical
  Review E}\ }\textbf {\bibinfo {volume} {101}},\ \bibinfo {pages} {013305}
  (\bibinfo {year} {2020})}\BibitemShut {NoStop}%
\bibitem [{\citenamefont {Brackbill}\ \emph {et~al.}(1992)\citenamefont
  {Brackbill}, \citenamefont {Kothe},\ and\ \citenamefont
  {Zemach}}]{brackbill1992continuum}%
  \BibitemOpen
  \bibfield  {author} {\bibinfo {author} {\bibfnamefont {J.~U.}\ \bibnamefont
  {Brackbill}}, \bibinfo {author} {\bibfnamefont {D.~B.}\ \bibnamefont
  {Kothe}},\ and\ \bibinfo {author} {\bibfnamefont {C.}~\bibnamefont
  {Zemach}},\ }\bibfield  {title} {\bibinfo {title} {A continuum method for
  modeling surface tension},\ }\href@noop {} {\bibfield  {journal} {\bibinfo
  {journal} {Journal of computational physics}\ }\textbf {\bibinfo {volume}
  {100}},\ \bibinfo {pages} {335} (\bibinfo {year} {1992})}\BibitemShut
  {NoStop}%
\bibitem [{\citenamefont {Jacqmin}(1999)}]{jacqmin1999calculation}%
  \BibitemOpen
  \bibfield  {author} {\bibinfo {author} {\bibfnamefont {D.}~\bibnamefont
  {Jacqmin}},\ }\bibfield  {title} {\bibinfo {title} {Calculation of two-phase
  navier--stokes flows using phase-field modeling},\ }\href@noop {} {\bibfield
  {journal} {\bibinfo  {journal} {Journal of Computational Physics}\ }\textbf
  {\bibinfo {volume} {155}},\ \bibinfo {pages} {96} (\bibinfo {year}
  {1999})}\BibitemShut {NoStop}%
\bibitem [{\citenamefont {Lafaurie}\ \emph {et~al.}(1994)\citenamefont
  {Lafaurie}, \citenamefont {Nardone}, \citenamefont {Scardovelli},
  \citenamefont {Zaleski},\ and\ \citenamefont
  {Zanetti}}]{lafaurie1994modelling}%
  \BibitemOpen
  \bibfield  {author} {\bibinfo {author} {\bibfnamefont {B.}~\bibnamefont
  {Lafaurie}}, \bibinfo {author} {\bibfnamefont {C.}~\bibnamefont {Nardone}},
  \bibinfo {author} {\bibfnamefont {R.}~\bibnamefont {Scardovelli}}, \bibinfo
  {author} {\bibfnamefont {S.}~\bibnamefont {Zaleski}},\ and\ \bibinfo {author}
  {\bibfnamefont {G.}~\bibnamefont {Zanetti}},\ }\bibfield  {title} {\bibinfo
  {title} {Modelling merging and fragmentation in multiphase flows with
  surfer},\ }\href@noop {} {\bibfield  {journal} {\bibinfo  {journal} {Journal
  of Computational Physics}\ }\textbf {\bibinfo {volume} {113}},\ \bibinfo
  {pages} {134} (\bibinfo {year} {1994})}\BibitemShut {NoStop}%
\bibitem [{\citenamefont {Lee}\ and\ \citenamefont
  {Fischer}(2006)}]{lee2006eliminating}%
  \BibitemOpen
  \bibfield  {author} {\bibinfo {author} {\bibfnamefont {T.}~\bibnamefont
  {Lee}}\ and\ \bibinfo {author} {\bibfnamefont {P.~F.}\ \bibnamefont
  {Fischer}},\ }\bibfield  {title} {\bibinfo {title} {Eliminating parasitic
  currents in the lattice boltzmann equation method for nonideal gases},\
  }\href@noop {} {\bibfield  {journal} {\bibinfo  {journal} {Physical Review
  E}\ }\textbf {\bibinfo {volume} {74}},\ \bibinfo {pages} {046709} (\bibinfo
  {year} {2006})}\BibitemShut {NoStop}%
\bibitem [{\citenamefont {Lee}\ and\ \citenamefont
  {Lin}(2003)}]{lee2003eulerian}%
  \BibitemOpen
  \bibfield  {author} {\bibinfo {author} {\bibfnamefont {T.}~\bibnamefont
  {Lee}}\ and\ \bibinfo {author} {\bibfnamefont {C.-L.}\ \bibnamefont {Lin}},\
  }\bibfield  {title} {\bibinfo {title} {An eulerian description of the
  streaming process in the lattice boltzmann equation},\ }\href@noop {}
  {\bibfield  {journal} {\bibinfo  {journal} {Journal of Computational
  Physics}\ }\textbf {\bibinfo {volume} {185}},\ \bibinfo {pages} {445}
  (\bibinfo {year} {2003})}\BibitemShut {NoStop}%
\bibitem [{\citenamefont {Lee}\ and\ \citenamefont
  {Lin}(2001)}]{lee2001characteristic}%
  \BibitemOpen
  \bibfield  {author} {\bibinfo {author} {\bibfnamefont {T.}~\bibnamefont
  {Lee}}\ and\ \bibinfo {author} {\bibfnamefont {C.-L.}\ \bibnamefont {Lin}},\
  }\bibfield  {title} {\bibinfo {title} {A characteristic galerkin method for
  discrete boltzmann equation},\ }\href@noop {} {\bibfield  {journal} {\bibinfo
   {journal} {Journal of Computational Physics}\ }\textbf {\bibinfo {volume}
  {171}},\ \bibinfo {pages} {336} (\bibinfo {year} {2001})}\BibitemShut
  {NoStop}%
\bibitem [{\citenamefont {Shan}\ and\ \citenamefont
  {Chen}(1993)}]{shan1993lattice}%
  \BibitemOpen
  \bibfield  {author} {\bibinfo {author} {\bibfnamefont {X.}~\bibnamefont
  {Shan}}\ and\ \bibinfo {author} {\bibfnamefont {H.}~\bibnamefont {Chen}},\
  }\bibfield  {title} {\bibinfo {title} {Lattice boltzmann model for simulating
  flows with multiple phases and components},\ }\href@noop {} {\bibfield
  {journal} {\bibinfo  {journal} {Physical review E}\ }\textbf {\bibinfo
  {volume} {47}},\ \bibinfo {pages} {1815} (\bibinfo {year}
  {1993})}\BibitemShut {NoStop}%
\bibitem [{\citenamefont {Li}\ \emph {et~al.}(2016)\citenamefont {Li},
  \citenamefont {Luo}, \citenamefont {Kang}, \citenamefont {He}, \citenamefont
  {Chen},\ and\ \citenamefont {Liu}}]{li2016lattice}%
  \BibitemOpen
  \bibfield  {author} {\bibinfo {author} {\bibfnamefont {Q.}~\bibnamefont
  {Li}}, \bibinfo {author} {\bibfnamefont {K.~H.}\ \bibnamefont {Luo}},
  \bibinfo {author} {\bibfnamefont {Q.}~\bibnamefont {Kang}}, \bibinfo {author}
  {\bibfnamefont {Y.}~\bibnamefont {He}}, \bibinfo {author} {\bibfnamefont
  {Q.}~\bibnamefont {Chen}},\ and\ \bibinfo {author} {\bibfnamefont
  {Q.}~\bibnamefont {Liu}},\ }\bibfield  {title} {\bibinfo {title} {Lattice
  boltzmann methods for multiphase flow and phase-change heat transfer},\
  }\href@noop {} {\bibfield  {journal} {\bibinfo  {journal} {Progress in Energy
  and Combustion Science}\ }\textbf {\bibinfo {volume} {52}},\ \bibinfo {pages}
  {62} (\bibinfo {year} {2016})}\BibitemShut {NoStop}%
\bibitem [{\citenamefont {Abadie}\ \emph {et~al.}(2015)\citenamefont {Abadie},
  \citenamefont {Aubin},\ and\ \citenamefont {Legendre}}]{abadie2015combined}%
  \BibitemOpen
  \bibfield  {author} {\bibinfo {author} {\bibfnamefont {T.}~\bibnamefont
  {Abadie}}, \bibinfo {author} {\bibfnamefont {J.}~\bibnamefont {Aubin}},\ and\
  \bibinfo {author} {\bibfnamefont {D.}~\bibnamefont {Legendre}},\ }\bibfield
  {title} {\bibinfo {title} {On the combined effects of surface tension force
  calculation and interface advection on spurious currents within volume of
  fluid and level set frameworks},\ }\href@noop {} {\bibfield  {journal}
  {\bibinfo  {journal} {Journal of Computational Physics}\ }\textbf {\bibinfo
  {volume} {297}},\ \bibinfo {pages} {611} (\bibinfo {year}
  {2015})}\BibitemShut {NoStop}%
\bibitem [{\citenamefont {Inamuro}\ \emph {et~al.}(2004)\citenamefont
  {Inamuro}, \citenamefont {Ogata}, \citenamefont {Tajima},\ and\ \citenamefont
  {Konishi}}]{inamuro2004lattice}%
  \BibitemOpen
  \bibfield  {author} {\bibinfo {author} {\bibfnamefont {T.}~\bibnamefont
  {Inamuro}}, \bibinfo {author} {\bibfnamefont {T.}~\bibnamefont {Ogata}},
  \bibinfo {author} {\bibfnamefont {S.}~\bibnamefont {Tajima}},\ and\ \bibinfo
  {author} {\bibfnamefont {N.}~\bibnamefont {Konishi}},\ }\bibfield  {title}
  {\bibinfo {title} {A lattice boltzmann method for incompressible two-phase
  flows with large density differences},\ }\href@noop {} {\bibfield  {journal}
  {\bibinfo  {journal} {Journal of Computational physics}\ }\textbf {\bibinfo
  {volume} {198}},\ \bibinfo {pages} {628} (\bibinfo {year}
  {2004})}\BibitemShut {NoStop}%
\bibitem [{\citenamefont {Zu}\ and\ \citenamefont {He}(2013)}]{zu2013phase}%
  \BibitemOpen
  \bibfield  {author} {\bibinfo {author} {\bibfnamefont {Y.}~\bibnamefont
  {Zu}}\ and\ \bibinfo {author} {\bibfnamefont {S.}~\bibnamefont {He}},\
  }\bibfield  {title} {\bibinfo {title} {Phase-field-based lattice boltzmann
  model for incompressible binary fluid systems with density and viscosity
  contrasts},\ }\href@noop {} {\bibfield  {journal} {\bibinfo  {journal}
  {Physical Review E}\ }\textbf {\bibinfo {volume} {87}},\ \bibinfo {pages}
  {043301} (\bibinfo {year} {2013})}\BibitemShut {NoStop}%
\bibitem [{\citenamefont {Kim}(2007)}]{kim2007phase}%
  \BibitemOpen
  \bibfield  {author} {\bibinfo {author} {\bibfnamefont {J.}~\bibnamefont
  {Kim}},\ }\bibfield  {title} {\bibinfo {title} {Phase field computations for
  ternary fluid flows},\ }\href@noop {} {\bibfield  {journal} {\bibinfo
  {journal} {Computer methods in applied mechanics and engineering}\ }\textbf
  {\bibinfo {volume} {196}},\ \bibinfo {pages} {4779} (\bibinfo {year}
  {2007})}\BibitemShut {NoStop}%
\bibitem [{\citenamefont {Lee}\ and\ \citenamefont
  {Kim}(2015)}]{lee2015efficient}%
  \BibitemOpen
  \bibfield  {author} {\bibinfo {author} {\bibfnamefont {H.~G.}\ \bibnamefont
  {Lee}}\ and\ \bibinfo {author} {\bibfnamefont {J.}~\bibnamefont {Kim}},\
  }\bibfield  {title} {\bibinfo {title} {An efficient numerical method for
  simulating multiphase flows using a diffuse interface model},\ }\href@noop {}
  {\bibfield  {journal} {\bibinfo  {journal} {Physica A: Statistical Mechanics
  and its Applications}\ }\textbf {\bibinfo {volume} {423}},\ \bibinfo {pages}
  {33} (\bibinfo {year} {2015})}\BibitemShut {NoStop}%
\bibitem [{\citenamefont {Lee}(2009)}]{lee2009effects}%
  \BibitemOpen
  \bibfield  {author} {\bibinfo {author} {\bibfnamefont {T.}~\bibnamefont
  {Lee}},\ }\bibfield  {title} {\bibinfo {title} {Effects of incompressibility
  on the elimination of parasitic currents in the lattice boltzmann equation
  method for binary fluids},\ }\href@noop {} {\bibfield  {journal} {\bibinfo
  {journal} {Computers \& Mathematics with Applications}\ }\textbf {\bibinfo
  {volume} {58}},\ \bibinfo {pages} {987} (\bibinfo {year} {2009})}\BibitemShut
  {NoStop}%
\bibitem [{\citenamefont {Fakhari}\ \emph {et~al.}(2017)\citenamefont
  {Fakhari}, \citenamefont {Mitchell}, \citenamefont {Leonardi},\ and\
  \citenamefont {Bolster}}]{fakhari2017improved}%
  \BibitemOpen
  \bibfield  {author} {\bibinfo {author} {\bibfnamefont {A.}~\bibnamefont
  {Fakhari}}, \bibinfo {author} {\bibfnamefont {T.}~\bibnamefont {Mitchell}},
  \bibinfo {author} {\bibfnamefont {C.}~\bibnamefont {Leonardi}},\ and\
  \bibinfo {author} {\bibfnamefont {D.}~\bibnamefont {Bolster}},\ }\bibfield
  {title} {\bibinfo {title} {Improved locality of the phase-field
  lattice-boltzmann model for immiscible fluids at high density ratios},\
  }\href@noop {} {\bibfield  {journal} {\bibinfo  {journal} {Physical Review
  E}\ }\textbf {\bibinfo {volume} {96}},\ \bibinfo {pages} {053301} (\bibinfo
  {year} {2017})}\BibitemShut {NoStop}%
\bibitem [{\citenamefont {Abu-Al-Saud}\ \emph {et~al.}(2018)\citenamefont
  {Abu-Al-Saud}, \citenamefont {Popinet},\ and\ \citenamefont
  {Tchelepi}}]{abu2018conservative}%
  \BibitemOpen
  \bibfield  {author} {\bibinfo {author} {\bibfnamefont {M.~O.}\ \bibnamefont
  {Abu-Al-Saud}}, \bibinfo {author} {\bibfnamefont {S.}~\bibnamefont
  {Popinet}},\ and\ \bibinfo {author} {\bibfnamefont {H.~A.}\ \bibnamefont
  {Tchelepi}},\ }\bibfield  {title} {\bibinfo {title} {A conservative and
  well-balanced surface tension model},\ }\href@noop {} {\bibfield  {journal}
  {\bibinfo  {journal} {Journal of Computational Physics}\ }\textbf {\bibinfo
  {volume} {371}},\ \bibinfo {pages} {896} (\bibinfo {year}
  {2018})}\BibitemShut {NoStop}%
\bibitem [{\citenamefont {Liang}\ \emph {et~al.}(2019)\citenamefont {Liang},
  \citenamefont {Li}, \citenamefont {Chen},\ and\ \citenamefont
  {Xu}}]{liang2019axisymmetric}%
  \BibitemOpen
  \bibfield  {author} {\bibinfo {author} {\bibfnamefont {H.}~\bibnamefont
  {Liang}}, \bibinfo {author} {\bibfnamefont {Y.}~\bibnamefont {Li}}, \bibinfo
  {author} {\bibfnamefont {J.}~\bibnamefont {Chen}},\ and\ \bibinfo {author}
  {\bibfnamefont {J.}~\bibnamefont {Xu}},\ }\bibfield  {title} {\bibinfo
  {title} {Axisymmetric lattice boltzmann model for multiphase flows with large
  density ratio},\ }\href@noop {} {\bibfield  {journal} {\bibinfo  {journal}
  {International Journal of Heat and Mass Transfer}\ }\textbf {\bibinfo
  {volume} {130}},\ \bibinfo {pages} {1189} (\bibinfo {year}
  {2019})}\BibitemShut {NoStop}%
\bibitem [{\citenamefont {Hua}\ and\ \citenamefont
  {Lou}(2007)}]{hua2007numerical}%
  \BibitemOpen
  \bibfield  {author} {\bibinfo {author} {\bibfnamefont {J.}~\bibnamefont
  {Hua}}\ and\ \bibinfo {author} {\bibfnamefont {J.}~\bibnamefont {Lou}},\
  }\bibfield  {title} {\bibinfo {title} {Numerical simulation of bubble rising
  in viscous liquid},\ }\href@noop {} {\bibfield  {journal} {\bibinfo
  {journal} {Journal of Computational Physics}\ }\textbf {\bibinfo {volume}
  {222}},\ \bibinfo {pages} {769} (\bibinfo {year} {2007})}\BibitemShut
  {NoStop}%
\bibitem [{\citenamefont {Popinet}(2018)}]{popinet2018numerical}%
  \BibitemOpen
  \bibfield  {author} {\bibinfo {author} {\bibfnamefont {S.}~\bibnamefont
  {Popinet}},\ }\bibfield  {title} {\bibinfo {title} {Numerical models of
  surface tension},\ }\href@noop {} {\bibfield  {journal} {\bibinfo  {journal}
  {Annual Review of Fluid Mechanics}\ }\textbf {\bibinfo {volume} {50}},\
  \bibinfo {pages} {49} (\bibinfo {year} {2018})}\BibitemShut {NoStop}%
\bibitem [{\citenamefont {Wang}\ \emph {et~al.}(2020)\citenamefont {Wang},
  \citenamefont {Semprebon}, \citenamefont {Liu}, \citenamefont {Zhang},\ and\
  \citenamefont {Kusumaatmaja}}]{wang2020modelling}%
  \BibitemOpen
  \bibfield  {author} {\bibinfo {author} {\bibfnamefont {N.}~\bibnamefont
  {Wang}}, \bibinfo {author} {\bibfnamefont {C.}~\bibnamefont {Semprebon}},
  \bibinfo {author} {\bibfnamefont {H.}~\bibnamefont {Liu}}, \bibinfo {author}
  {\bibfnamefont {C.}~\bibnamefont {Zhang}},\ and\ \bibinfo {author}
  {\bibfnamefont {H.}~\bibnamefont {Kusumaatmaja}},\ }\bibfield  {title}
  {\bibinfo {title} {Modelling double emulsion formation in planar
  flow-focusing microchannels},\ }\href@noop {} {\bibfield  {journal} {\bibinfo
   {journal} {Journal of Fluid Mechanics}\ }\textbf {\bibinfo {volume} {895}}
  (\bibinfo {year} {2020})}\BibitemShut {NoStop}%
\bibitem [{\citenamefont {Hysing}\ \emph {et~al.}(2009)\citenamefont {Hysing},
  \citenamefont {Turek}, \citenamefont {Kuzmin}, \citenamefont {Parolini},
  \citenamefont {Burman}, \citenamefont {Ganesan},\ and\ \citenamefont
  {Tobiska}}]{hysing2009quantitative}%
  \BibitemOpen
  \bibfield  {author} {\bibinfo {author} {\bibfnamefont {S.-R.}\ \bibnamefont
  {Hysing}}, \bibinfo {author} {\bibfnamefont {S.}~\bibnamefont {Turek}},
  \bibinfo {author} {\bibfnamefont {D.}~\bibnamefont {Kuzmin}}, \bibinfo
  {author} {\bibfnamefont {N.}~\bibnamefont {Parolini}}, \bibinfo {author}
  {\bibfnamefont {E.}~\bibnamefont {Burman}}, \bibinfo {author} {\bibfnamefont
  {S.}~\bibnamefont {Ganesan}},\ and\ \bibinfo {author} {\bibfnamefont
  {L.}~\bibnamefont {Tobiska}},\ }\bibfield  {title} {\bibinfo {title}
  {Quantitative benchmark computations of two-dimensional bubble dynamics},\
  }\href@noop {} {\bibfield  {journal} {\bibinfo  {journal} {International
  Journal for Numerical Methods in Fluids}\ }\textbf {\bibinfo {volume} {60}},\
  \bibinfo {pages} {1259} (\bibinfo {year} {2009})}\BibitemShut {NoStop}%
\end{thebibliography}%

\end{document}